***Chapter 1***

# $CaCu_3Ti_4O_{12}$ (CCTO) CERAMICS FOR CAPACITOR APPLICATIONS

***Rainer Schmidt*** **and** ***Derek C. Sinclair*****[#]**

## ABSTRACT

$CaCu_3Ti_4O_{12}$ (CCTO) ceramics are potential candidates for capacitor applications due to their large dielectric permittivity ($\varepsilon'$) values of up to 300 000. The underlying mechanism for the high $\varepsilon'$ is an internal barrier layer capacitor (IBLC) structure of insulating grain boundaries (GB) and conducting grain interiors (bulk). This behaviour is reviewed and discussed in detail. The origin of the IBLC structure is attributed to a small Cu non-stoichiometry in nominally insulating $CaCu_3Ti_4O_{12}$, which varies between the GBs and bulk. Such non-stoichiometry effects are studied in detail by analyzing bulk ceramics of different composition within the ternary CaO-CuO-$TiO_2$ phase diagram using X-ray diffraction (XRD), scanning electron microscopy (SEM) and impedance spectroscopy (IS). At least two defect mechanisms are suggested to exist. It is further shown that the development of the defect mechanisms in CCTO and the concomitant formation of the IBLC structure strongly depend on the processing conditions of CCTO ceramic pellets such as the sintering temperature. Nominally stoichiometric CCTO bulk ceramics sintered at different temperatures are analyzed using XRD, SEM and IS. The performance of CCTO ceramics for IBLC applications is controlled by subtle modifications in the compound stoichiometry that is strongly dependent on the ceramic sintering temperature.

[*] Rainer Schmidt: (a) Universidad Complutense de Madrid, Departamento Física Aplicada III, GFMC, Facultad de Ciencias Físicas, 28040 Madrid, Spain. (b) The University of Sheffield, Department of Materials Science and Engineering, Mappin Street, Sheffield S1 3JD, United Kingdom.
Corresponding author. Electronic Mail: rainerxschmidt@googlemail.com
[#] Derek C. Sinclair: The University of Sheffield, Department of Materials Science and Engineering, Mappin Street, Sheffield S1 3JD, United Kingdom.

# I. INTRODUCTION

The ternary oxide compound $CaCu_3Ti_4O_{12}$ (CCTO) has been considered for applications as a capacitor material due to its exceptionally high dielectric permittivity values. In polycrystalline ceramics the highest values reported are ≈ 300 000, ≈ 700 for thin films and ≈ 100 000 for single crystals [1-12].

It is now relatively well established that such giant permittivity values are extrinsic in origin. In single crystals high permittivity has been interpreted as an extrinsic electrode – sample interface barrier [13, 14], whereas in polycrystalline CCTO ceramics it is commonly explained in terms of an internal barrier layer capacitor (IBLC) structure of thin insulating grain boundaries (GBs) and conducting grain interior (bulk) regions [15, 16]. Electrode–sample interface barriers can also contribute to the high permittivity in ceramics [17], which has been interpreted previously as a surface barrier layer structure (SBLC) [18]. Since GB regions are generally expected to cover the conducting grain interiors, IBLCs may in fact resemble a core-shell microstructure which is well known to exhibit large capacitance [19].

Conventional core-shell devices generally require sophisticated processing to fully cover conducting grains or an equivalent interior phase with an insulating secondary phase. Contrarily, the IBLC core-shell microstructure in CCTO may arise directly from subtle chemical differences between the CCTO phases being formed at GB and grain interior (bulk) regions during ceramic processing and a secondary non-CCTO GB phase is not present. However, the stoichiometric differences between GBs and bulk in CCTO are not yet fully understood despite considerable research efforts in recent years. The work presented here is therefore intended to shed light on the defect chemistry and non-stoichiometry of CCTO.

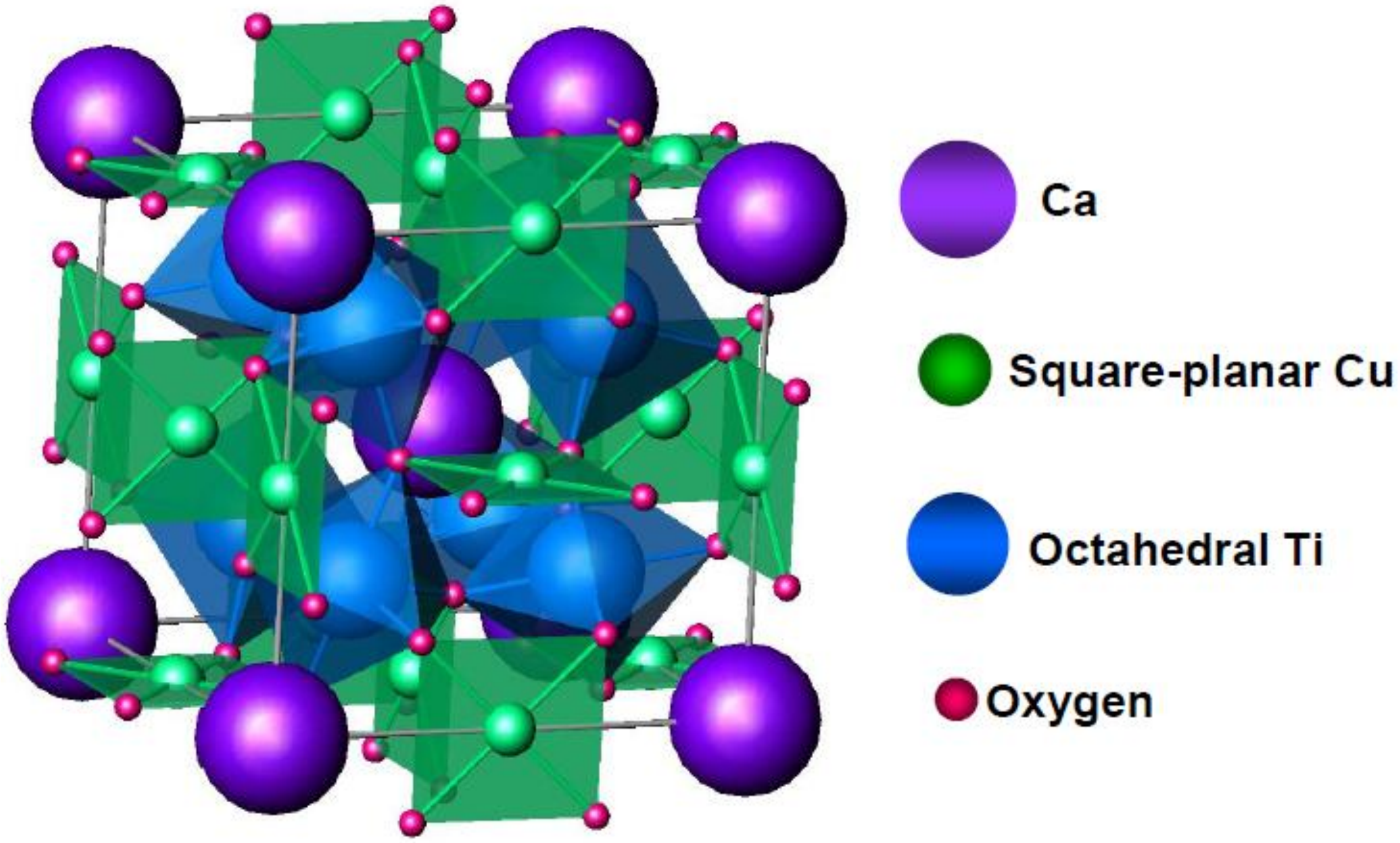

**Figure 1.** $CaCu_3Ti_4O_{12}$: A strongly distorted, A-site ordered double-perovskite structure.

In particular, the influence of processing conditions on the defect mechanisms and the formation of the IBLC structure is reported. This work comprises two experimental parts, where (A) ceramics with various compositions in the CaO-CuO-$TiO_2$ ternary phase diagram and (B) nominal CCTO ceramics sintered at various temperatures have been comprehensively investigated and characterised by a variety of techniques.

The crystal structure of CCTO is a 1:3 A-site ordered perovskite ($A'A''_3B_4O_{12}$). The oxygen octahedra are heavily tilted and the A" site Cu cations adopt a four-fold square-planar coordination. The unit cell is a doubled simple perovskite cell (Figure 1) and is commonly indexed as cubic Im-3. The octahedral tilting occurs because of the large cation size mismatch between the A'-site $Ca^{2+}$ (1.34 Å) and the A"-site $Cu^{2+}$ (0.57 Å) cations [20].

# II. Impedance Spectroscopy (IS) and Non-Stoichiometry Models in CCTO

## A. Basic Principles of Impedance Spectroscopy

Temperature dependent impedance spectroscopy (IS) enables different contributions to the dielectric and resistive properties of condensed matter to be deconvoluted and characterized separately [21, 22]. IS is therefore the method of choice to separately determine and investigate the dielectric properties of GB and grain interior (bulk) regions in electrically inhomogeneous electroceramics such as CCTO [4]. IS experiments consist of an electric stimulus in terms of a time ($t$)-dependent alternating voltage signal $U$ of angular frequency $\omega$ and amplitude $U_0$ applied to a sample, and effectively the amplitude $I_0$ and phase shift $\delta$ of the current response signal $I$ are measured.

$$U(\omega, t) = \mathbf{U_0} \cos(\boldsymbol{\omega} t); \qquad \rightarrow \qquad I(\omega, t) = I_0 \cos(\boldsymbol{\omega} t - \delta); \tag{1}$$

All parameters defined by the applied voltage signal ($U_0$, $\omega$) are printed in blue/bold, the measured parameters of the current response are in red ($I_0$, $\delta$). One phase of the applied voltage signal corresponds to a 2π (360°) rotation of the voltage arrow $U(\omega, t)$ shown on the phasor diagram in Figure 2. The current response of ideal circuit elements is: (1) in-phase with the applied voltage in the case of an ideal resistor $R$; (2) out-of-phase by $\delta = -\pi/2$ for an ideal capacitor $C$; and (3) out-of-phase by $\delta = +\pi/2$ for an ideal inductor. All phase angles are time independent and all arrows rotate at constant angles for a given frequency.

The impedance can therefore be defined as a time-independent complex number $Z^*$ (= $Z'$ + i$Z''$), where the impedance fraction in-phase with $U(\omega,t)$ is defined as the real part, and the fraction +/- π/2 out-of-phase as the positive/negative imaginary part. The complex impedance $Z^*$ definitions for several electronic circuit elements are given in Figure 2.

It is well established that different dielectric relaxation processes detected by IS such as those found at the GB and bulk regions can, in the simplest case, be understood in terms of the brick work layer model where each GB or bulk type contribution is described by one RC circuit element consisting of a resistor and capacitor in parallel [23]. Several dielectric relaxations in series can be described by several RC elements connected in series, where the macroscopic impedance is just the sum of all series RC impedances. The RC model works particularly well for insulators such as dielectrics, where the capacitor describes the ability of the material to store charge and the parallel resistor describes the leakage current due to some un-trapped charge carriers bypassing the ideal charge storage element.

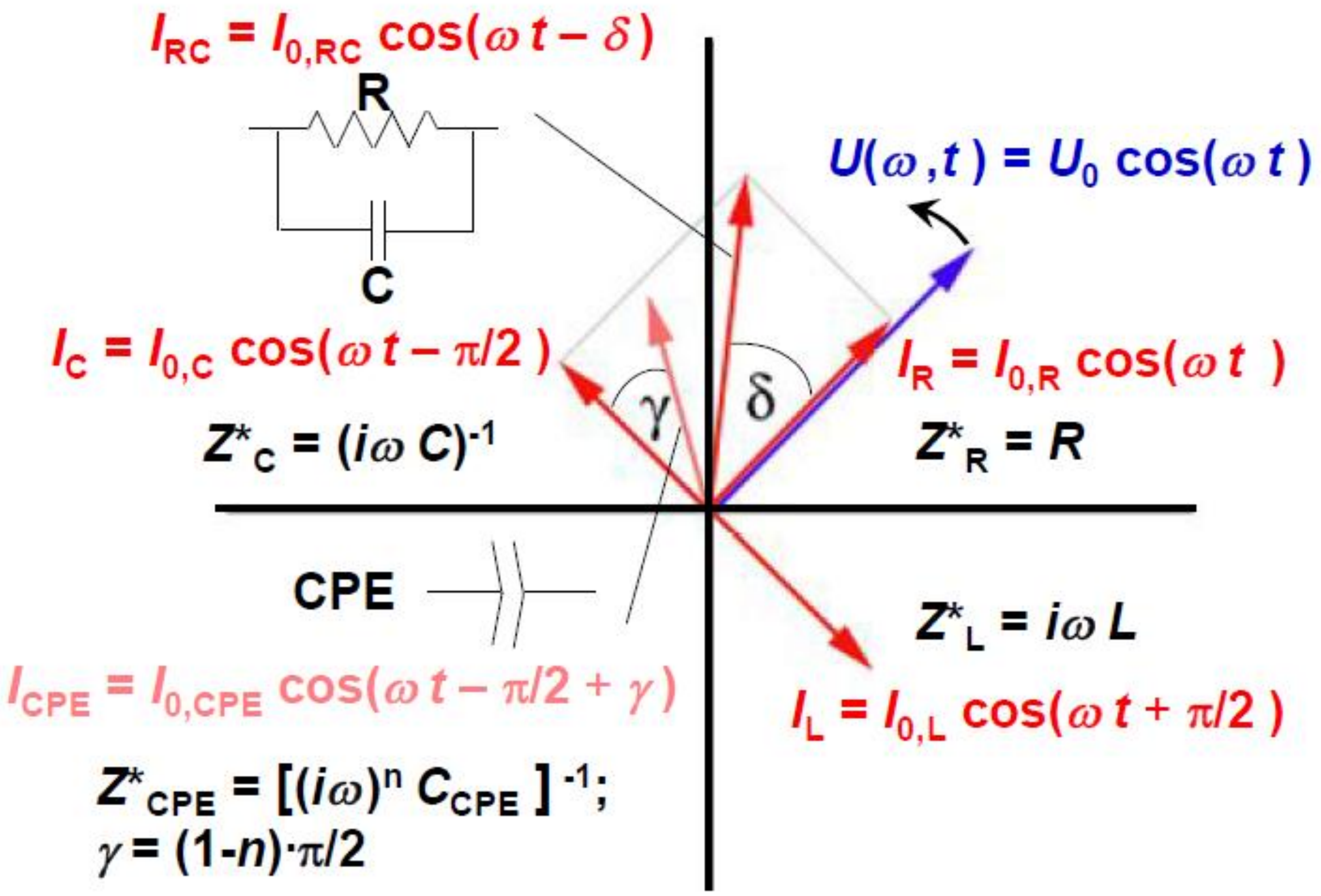

**Figure 2.** Impedance response of circuit elements on a phasor diagram: applied voltage stimulus $U(\omega t)$, current response $I_R$ from an ideal resistor, $I_C$ from an ideal capacitor, $I_{RC}$ from an ideal parallel resistor-capacitor (RC) element with phase shift $\delta$, $I_L$ from an ideal inductor, and $I_{CPE}$ from a constant phase element (CPE) for a non-ideal capacitor with a frequency independent phase shift $\gamma$ with respect to $I_C$.

In reality, the dielectric relaxation processes in electroceramics like CCTO are usually not ideal and, therefore, experimental IS data can usually not be modeled by ideal RC elements. To fit the data to an adequate equivalent circuit model the non-ideality in one specific dielectric relaxation can be accounted for in the respective RC element by connecting a constant-phase element (CPE) in parallel (R-CPE-C) or by replacing the ideal capacitor with a CPE (R-CPE). The phase angle and the complex impedance $Z_{CPE}$* of a CPE are shown in Figure 2, where $C_{CPE}$ is the CPE specific capacitance. The CPE capacitance obtained from fits to an R-CPE circuit can be converted into a real capacitance given in [Farad] using a standard procedure [24].

The critical exponent $n$ defined in Figure 2 has typical values of $n = 0.6 – 1$. $n = 1$ constitutes the ideal case of an ideal capacitor for an ideal dielectric relaxation. In a non-ideal R-CPE-C or R-CPE circuit decreasing $n$ values indicate a broadening of the respective dielectric relaxation peak as a reflection of the broadening of the distribution of relaxation times, $\tau$, across the sample. In an ideal RC element $\tau$ is given by $\tau$ = RC. The exact shape of the distribution of $\tau$ is complicated to be determined from IS data, and the exponent $n$ constitutes a semi-empirical parameter to reflect an increasing width of the distribution in $\tau$ by decreasing $n$ values.

In addition to the complex impedance $Z'$-$Z''$, IS data can also be represented in alternative formats using the standard conversions shown in Figure 3 for the real part of the dielectric permittivity ($\varepsilon'$) or capacitance ($C'$) and the imaginary part of the dielectric modulus ($M''$). Such plots often contain important information on the resistance or permittivity of a specific relaxation. In a scenario of a ceramic sample such as CCTO with one GB and one bulk (b) relaxation as represented by two RC elements connected in series the following spectral features are expected as shown from the simulations presented in Figure 3:

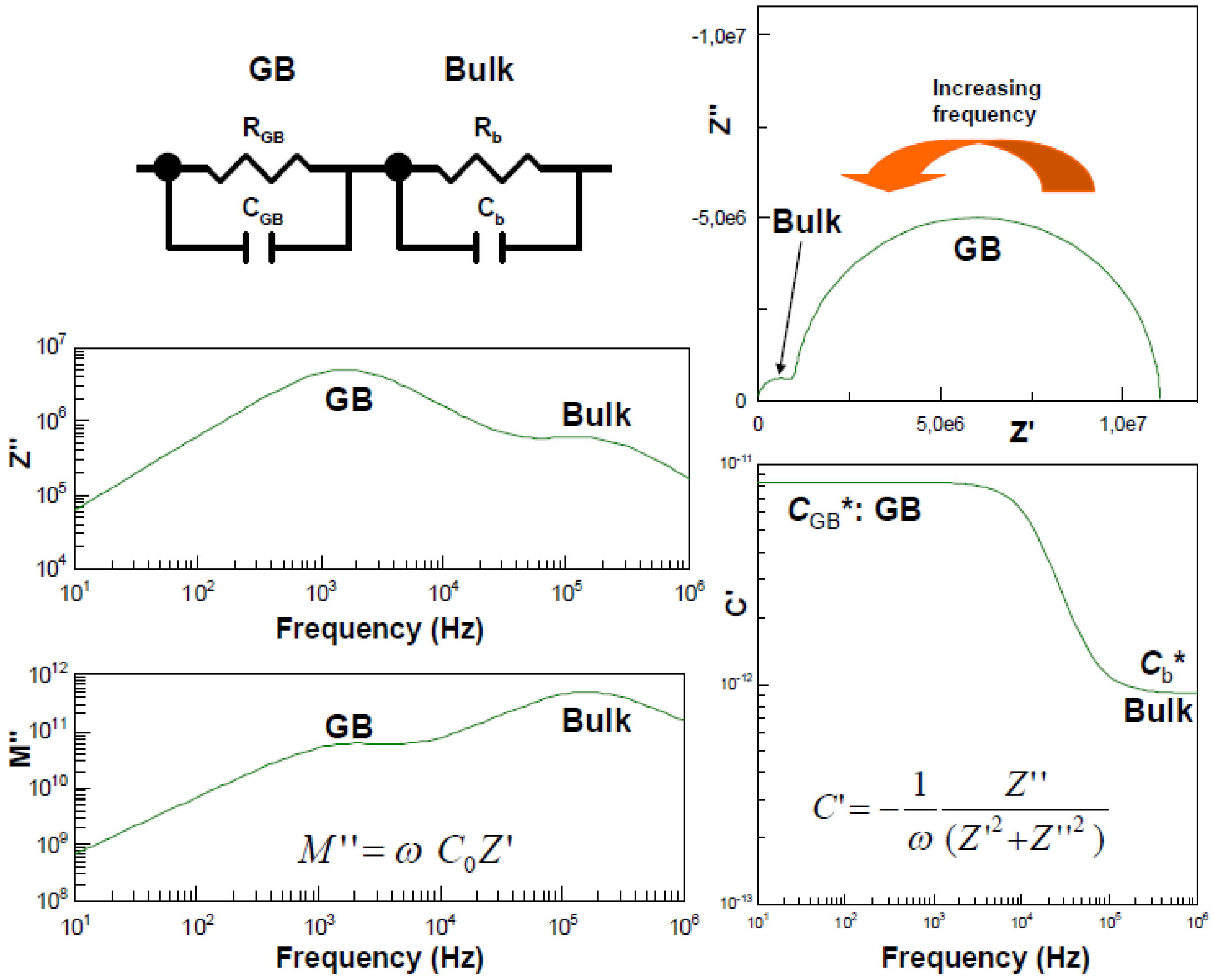

**Figure 3.** Simulated IS data for two RC elements connected in series, presented in different formats: $Z''$ [Ω] vs frequency ($f$), $M''$ vs $f$, $Z''$ [Ω] vs $Z'$ [Ω] and $C'$ [Farad] vs $f$. Simulations were carried out with $R_{GB}$ = 10 MΩ, $R_b$ = 1 MΩ, $C_{GB}$ = 10 pF and $C_b$ = 1 pF.

(1.) In plots of $Z''$ vs $Z'$ two semicircles appear with each semicircle diameter corresponding to the respective resistance $R$, (2.) $Z''$ vs frequency ($f$) plots show two relaxation peaks with peak heights of $R/2$ for each relaxation, (3.) $C'$ vs $f$ or $\varepsilon'$ vs $f$ display two approximately $f$ independent $C'$ ($\varepsilon'$) plateaus $C_{GB}$* and $C_b$*, that correspond approximately to the GB ($C_{GB}$) and bulk capacitance ($C_b$) and show an intermediate sharp drop at a distinct $f$, (4.) $\varepsilon''$ vs $\varepsilon'$ plots display one semicircle with the diameter being directly proportional to the GB permittivity (not shown), and (5.) $M''$ vs $f$ displays two relaxation peaks with the peak heights being inversely proportional to each capacitance, $1/C$.

The resistance and capacitance values used for the simulations are given in the Figure 3 caption, where the values were chosen to represent a realistic scenario of one extrinsic GB- and one intrinsic bulk-type relaxation.

In the framework of the brick work layer model of two RC elements connected in series, the GB and bulk capacitance plateaus $C_{GB}$* and $C_b$* depicted in the $C'$ vs $f$ plot in Figure 3 are both only an approximation for the GB and bulk capacitance $C_{GB}$ and $C_b$ respectively. In fact, both plateaus are composite terms, where $C_{GB}$* contains all resistor and capacitor terms and $C_b$* contains a contribution from the GB capacitance. This is apparent from Figure 3, since $C_{GB}$* and $C_b$* do not coincide exactly with the values of the capacitors $C_{GB}$ (10 pF) and $C_b$ (1 pF). The exact expressions for $C_{GB}$* and $C_b$* are given by equations (2) and (3) respectively.

$$C_{GB}^{*} = \frac{R_{GB}^{2} C_{GB} + R_{b}^{2} C_{b}}{\left(R_{GB} + R_{b}\right)^{2}} \tag{2}$$

$$C_{b}^{*} = \frac{C_{GB} \times C_{b}}{C_{GB} + C_{b}}; \tag{3}$$

In the dielectric permittivity format of $\varepsilon_{GB}$*, $\varepsilon_{b}$*, $\varepsilon_{GB}$ and $\varepsilon_{b}$, equations (2) and (3) are equivalent. In the case that the GB and bulk resistance $R_{GB}$ and $R_{b}$ are sufficiently different, i.e. $R_{GB}$ » $R_{b}$, then $C_{GB}$* constitutes an excellent estimate for $C_{GB}$. This is usually the case for CCTO where $R_{GB}$ and $R_{b}$ commonly vary by more than 3 orders of magnitude. $C_{b}$* is a good estimate for $C_{b}$ for the case where the two capacitors $C_{GB}$ and $C_{b}$ are sufficiently different, i.e. $C_{GB}$ » $C_{b}$. This is not always the case for CCTO as is discussed below in section IV./A.

In CCTO with distinctively different $R_{GB}$ and $R_{b}$ values the *M*'' vs *f* plots are particularly useful to display the bulk relaxation peak. *Z*'' vs *f* plots highlight the relaxation peak with the highest resistance *R* (e.g. the GB peak in CCTO), whereas the *M*'' vs *f* plots highlight the smallest capacitance (i.e. the bulk peak in CCTO).

The approximate peak frequencies for GB or bulk relaxation peaks as depicted in Figure 3 are given in equation 4 for the respective plots of -*Z*'' vs *f* (GB) and *M*'' vs *f* (bulk), using two ideal RC elements connected in series. Equation (4) implies an extrinsic GB-type relaxation with large resistance and capacitance to appear at lower *f* than the bulk, which is a common feature in experimental impedance spectra from electroceramics.

The different ordinates of the GB and bulk peaks are indicated in equation (5) for a CCTO scenario with distinct differences between the GB and bulk RC elements.

$$f_{\max}(-Z'') \approx \frac{1}{2\pi\ R_{GB} C_{GB}};\quad f_{\max}(M'') \approx \frac{1}{2\pi\ R_{b} C_{b}} \tag{4}$$

$$-Z''(f_{\max}) \approx \frac{R_{GB}}{2};\quad M''(f_{\max}) \approx \frac{C_{0}}{2C_{b}} \tag{5}$$

$C_{0}$ represents the capacitance of the empty measuring cell in vacuum. The differences in the peak ordinates in equation (5) guarantee that in -*Z*'' vs *f* the relaxation peaks representing the largest resistance (GB) and in *M*'' vs *f* the peaks representing the smallest capacitance (bulk) are the most strongly pronounced (see Figure 3).

In the case where both resistors in Figure 3 are thermally activated and follow Arrhenius type behavior, as is the case for CCTO, the dielectric relaxation peaks and the drop in *C*' depicted in Figure 3 would all move to higher *f* with increasing temperature, *T*. This appears as if the full spectrum shifts to higher *f* as is demonstrated below in Figure 4b for the example of a *C*' vs *f* plot.

## B. Impedance Spectroscopy on CCTO

The typical ceramic IBLC structure for CCTO ceramics is depicted in Figure 4a, where the semiconducting grain interiors (bulk) are shown in dark/blue whereas the surrounding insulating GB regions are colourless. No percolation path for high conductivity is available and the macroscopic direct current (dc) resistance is insulating. By carrying out IS it appears that at sufficiently high *f* a sharp drop occurs in *C*' vs *f* plots and the giant extrinsic GB permittivity reduces to the bulk relative dielectric permittivity of ≈ 100 (Figure 4b) [18, 25].

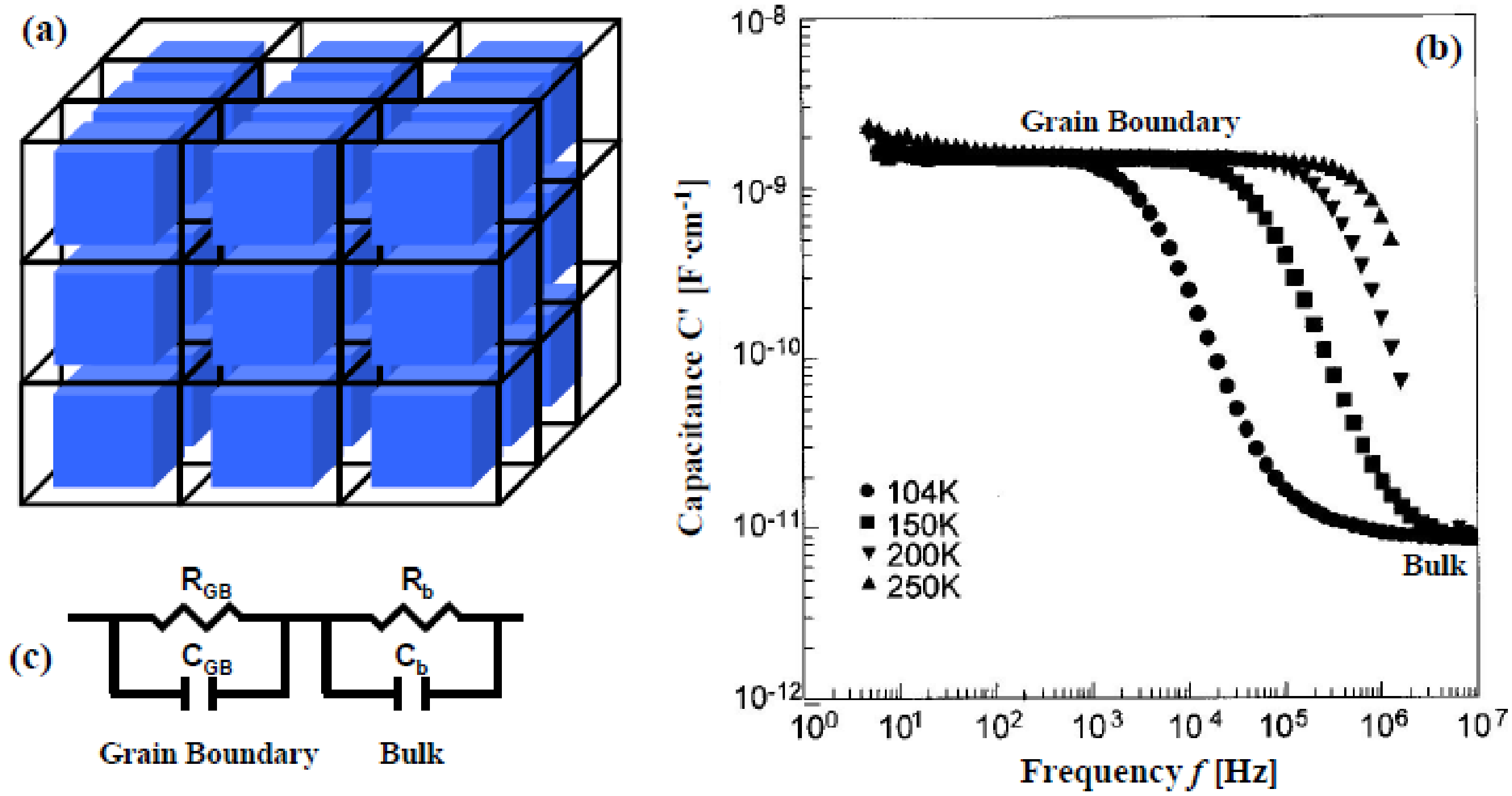

**Figure 4.** Internal barrier layer capacitor (IBLC) structure for CCTO: **(a)** Ceramic grains are represented schematically by cubes (solid lines). The semiconducting (grain) bulk regions are represented by smaller (blue) cubes. The surrounding insulating grain boundary areas are colourless. **(b)** Impedance spectroscopy (IS) data from CCTO ceramics represented in the notation of the real part of capacitance *C*' vs *f*. Image reproduced from [15], with permission from the American Institute of Physics (AIP). **(c)** Idealized equivalent circuit model to account for GB and bulk (b) dielectric relaxation processes.

This is the typical behaviour of an IBLC structure, where the sharp *C*' drop occurs when the mean electron conduction path permitted by the applied alternating voltage signal decreases below the average grain size at increased *f*. The low-*f* dielectric response is therefore dominated by the insulating GB and the high-*f* response by the conducting bulk contribution. Both dielectric relaxation processes (GB and bulk) are represented in an idealised way by two RC elements connected in series (Figure 4c). In CCTO ceramics the inhomogeneous dielectric behaviour can be pronounced: GB and bulk resistance can differ by a factor of up to $\approx 10^5$. This is demonstrated in Figure 5, where the GB and bulk (b) semicircles are not both visible at the same temperature. Only at high *T* (300 K) does the GB exhibit sufficiently low resistance to be fully developed as an arc in the *Z*" vs *Z*' plot, whereas the bulk resistance is small and the bulk relaxation peak occurs at a higher *f* than the measured spectrum. Only a non-zero intercept of the data with the *Z*' axis appears at the upper limit of *f* as demonstrated by the (red) bulk (b) arrow in the Figure 5a inset.

At low *T* (104, 115 and 130 K) the bulk resistance is larger, the bulk relaxation peak has shifted to lower *f* in agreement with equation (4) and can now be clearly resolved. On the other hand, the GB resistance is now too large to be resolved in the form of a fully developed GB semicircle and only the onset is visible.

The decreasing size of the bulk semicircle in Figure 5(b) with increasing *T* is indicative of thermally activated charge transport. The associated charge transport activation energies, $E_a$, associated with either GB or bulk CCTO conduction processes, have been reported to differ up to a factor of ≈ 10 [16, 26]. This is highlighted in Figure 6, where $R_{GB}$ and $R_b$ values were determined for a CCTO ceramic sample from the respective semicircle diameters in *Z*" vs *Z*' plots.

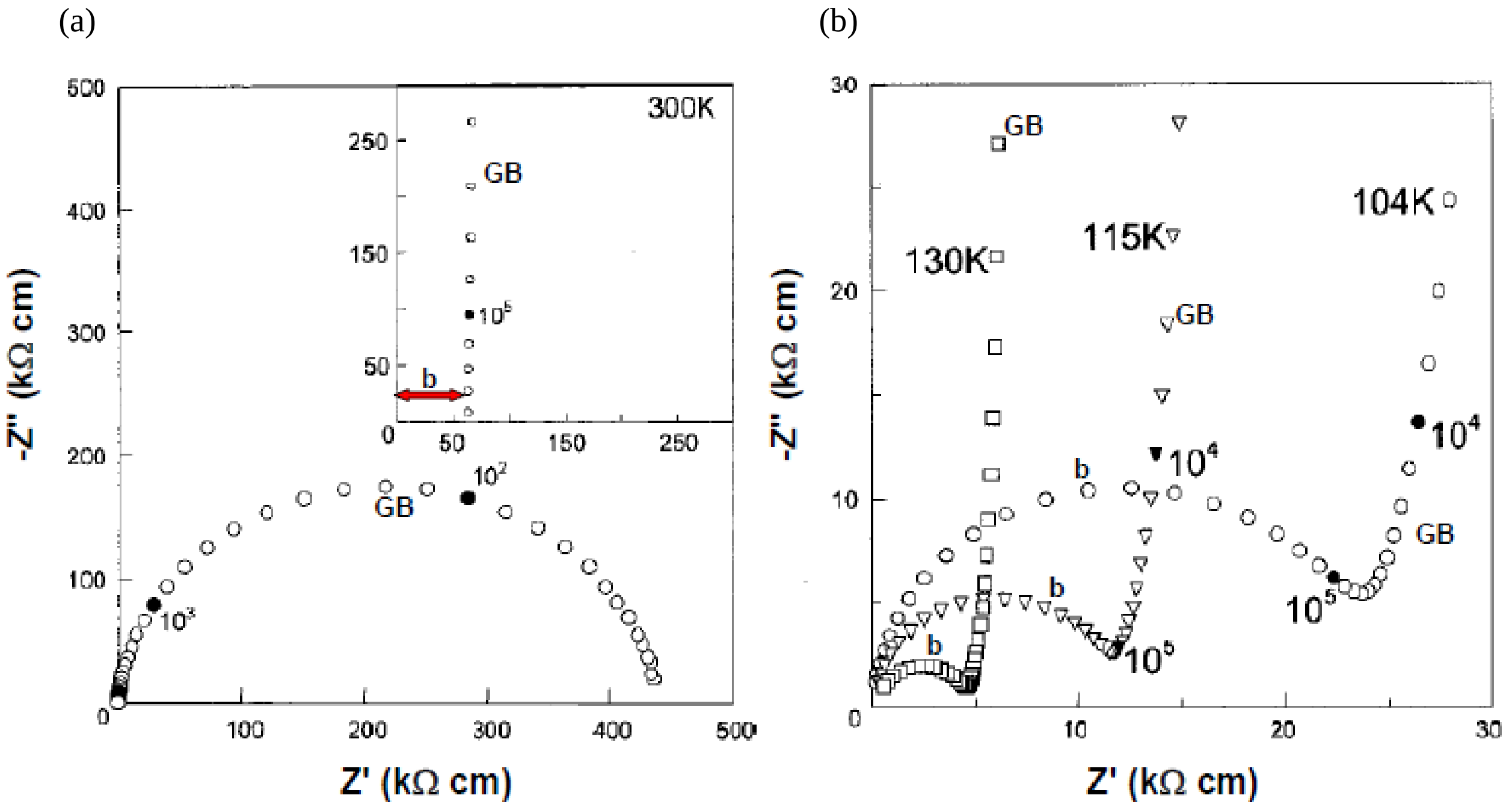

**Figure 5.** Complex impedance *Z*'' vs *Z*' plots for ceramic CCTO. GB and bulk (b) contributions are indicated. **(a)** At 300 K only the GB semicircle is visible, where the bulk (b) only appears as a non-zero intercept of the data curve with the *Z*' axis at high *f* (see bulk arrow (b) in the Figure inset). **(b)** On cooling the bulk relaxation peak shifts to lower *f* and is visible at 104, 115 and 130 K. The change in semicircle diameter indicates thermally activated charge transport. The GB semicircle is only partially visible due to high *R* and cannot be resolved. Images reproduced from [15] with permission from the American Institute of Physics (AIP).

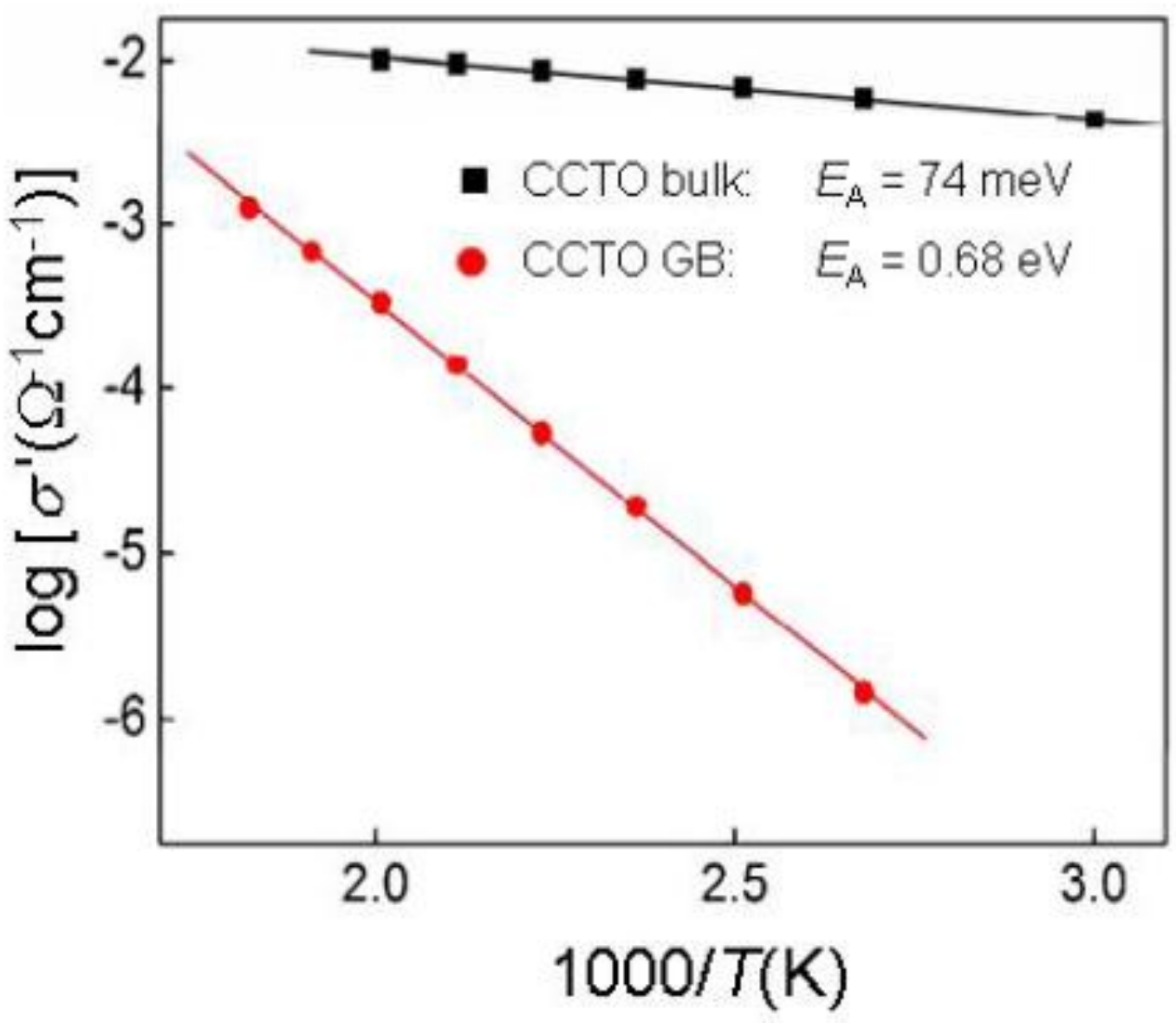

**Figure 6.** Arrhenius plots of GB and bulk conductivities for CCTO. The GB and bulk activation energies $E_A$ as well as the nominal conductivity values differ significantly. Image reproduced from [27], with permission from M. Li.

The data are plotted in the conductivity format of $1/R_{GB}$ and $1/R_b$. The large difference displayed in $E_a$ as well as in the nominal resistance between $R_{GB}$ and $R_b$ is a strong indicator that different charge transport mechanisms must exist. This in turn implies that also perceptible chemical differences between GB and bulk regions must exist to cause this variation. These chemical differences between GB and bulk CCTO are not well understood, especially the low bulk resistance in nominally stoichiometric (and therefore insulating) $CaCu_3Ti_4O_{12}$. This has led to controversial discussions in the literature over potential defect mechanisms and the non-stoichiometry that are responsible for inducing semiconductivity in the grains.

## C. Non-Stoichiometry Models in CCTO

In the literature several controversial reports exist, where different defect mechanisms and associated solid solutions have been proposed to explain CCTO bulk semi-conductivity and the IBLC structure, reporting on various cation distributions and different types of bulk charge transport (n-type electron or p-type hole conduction). Such conduction may rely on electron transport based on mixed valence $Ti^{3+}/Ti^{4+}$ or $Cu^{1+}/Cu^{2+}$ (n-type), or $Cu^{2+}/Cu^{3+}$ (p-type). The following defect mechanisms, solid solutions and possible compensation mechanisms have been reported in the literature [28-30] or appear to be plausible from the results presented here; brackets ( ) indicate A" Cu sites:

(A) $O_o^{2-} + 2Ti^{4+} \rightarrow V_0^{''} + 2Ti^{3+} + 1/2O_2 \uparrow$ : $Ca(Cu_3)Ti_{4-u}^{4+}Ti_u^{3+}O_{12-u/2}$

(B) $3Cu^{2+} \rightarrow 2Cu^{1+} + Ti_{Cu}^{4+} + Cu \uparrow$ : $Ca(Cu_{3-3v}^{2+}Cu_{2v}^{+}Ti_v^{4+})Ti_4O_{12}$

(C) $Cu^{+} + Ti^{4+} \rightarrow Cu^{2+} + Ti^{3+}$ : $Ca(Cu_{3-v}^{2+}Ti_v^{4+})Ti_{4-2v}^{4+}Ti_{2v}^{3+}O_{12}$

(D) $3Cu^{2+} \rightarrow V_{Cu}^{''} + 2Cu^{3+} + Cu \uparrow$ : $Ca(Cu_{3-3y}^{2+}Cu_{2y}^{3+}V_y^{''})Ti_4O_{12}$

(E) $O_o^{2-} + Ti^{4+} \rightarrow V_0^{''} + Cu_{Ti}^{2+} + 1/2O_2 \uparrow$ : $Ca(Cu_3^{2+})Ti_{4-w}^{4+}Cu_w^{2+}O_{12-w}$

(A) represents oxygen loss compensated by $Ti^{4+}$ reduction. (B) represents Cu loss or deficiency compensated by partial reduction of $Cu^{2+}$ to $Cu^{1+}$ and partial occupation of Cu sites by $Ti^{4+}$. Processes (A) and (B) would most likely take place during heat treatment at elevated temperatures. In (C) the $Cu^{2+}$ reduction at elevated temperatures described in (B) is now assumed to be fully reversible upon cooling by reoxidation, whereas the Cu deficiency persists. The reoxidation of $Cu^{1+}$ to $Cu^{2+}$ in turn is compensated by partial reduction of $Ti^{4+}$ to $Ti^{3+}$; i.e. an internal redox process takes place. Only the redox process is displayed in (C), the high temperature Cu reduction is identical to (B). (D) represents Cu loss due to Cu segregation or volatilisation and compensation is via partial oxidation of $Cu^{2+}$ to $Cu^{3+}$. (E) corresponds to Cu excess and compensation is by $Cu^{2+}$ occupying Ti sites and the formation of oxygen vacancies. Alternatively, compensation in (E) could also occur via $Cu^{2+}$ oxidation to $Cu^{3+}$ and fewer oxygen vacancies would form. Models (A), (B) and (C) are supported by findings of n-type conduction [29]. Process (B) is consistent with reports of $Cu^{1+}$ from XPS studies [31]. (D) is supported by findings of (I.) Cu loss [30, 32]; (II.) the presence of $Cu^{3+}$ from XPS studies [30], and (III.) p-type bulk semiconductivity determined from Hall-effect measurements [8]. (E) has not been mentioned in the literature but may be applicable as discussed below.

## D. Doped CCTO

One or more of the abovementioned defect mechanisms (A - E) must be present in CCTO ceramics with an IBLC structure to explain chemical differences between insulating GBs and semi-conducting bulk. In the literature a large amount of publications can be found to demonstrate that chemical doping can induce new or possibly modify one or more of the abovementioned defect mechanisms leading to distinctively varied IBLC structures with modified resistance and permittivity values of GB and grain interior regions. In Figure 7 two examples are given: A'-site Sr-doping in $Ca_{1-x}Sr_xCu_3Ti_4O_{12}$ [5] and A''-site Mn doping in $CaCu_{3-x}Mn_xTi_4O_{12}$ [28]. A'-site Ca substitution by Sr leads to a modest increase in GB and bulk capacitance, whereas A''-site Cu substitution by Mn has a much more dramatic effect: on the right hand panel of Figure 7 the dielectric inhomogeneity of the ceramic is shown to disappear, whereby the bulk resistance increases [33]. The $CaCu_{2.94}Mn_{0.06}Ti_4O_{12}$ ceramic is homogeneous, only the bulk capacitance is visible, the grains are highly resistive [27] and the defect mechanism leading to semiconductivity in undoped CCTO is suppressed or absent. This may lead to significant conclusions:

If the CCTO bulk semi-conductivity can be "killed" by Mn-doping, this is a clear indication that the differences between GB and bulk regions in CCTO may well emerge from subtle chemical differences in a GB-type and a bulk-type CCTO phase and not from a secondary non-CCTO phase. Such a potential secondary non-CCTO phase needed to be a GB-phase with a small volume fraction, because the CCTO bulk is commonly XRD phase-pure. The non-CCTO GB phase would be expected to fully cover the semi-conducting grain interior regions and not permit a conducting path across a ceramic sample. Such a phase has never been observed unambiguously and is in fact not plausible regarding the effects of Mn-doping. From Figure 7 it should be noted that the step-like decrease in *C*' vs *f* (left panel) and the step-like increase in *C*' vs *T* (right panel) for undoped CCTO may be somewhat opposing trends, but they are both fully consistent with a circuit dominated by two RC elements connected in series for GBs and bulk, as was pointed out in a recent simulation study [34].

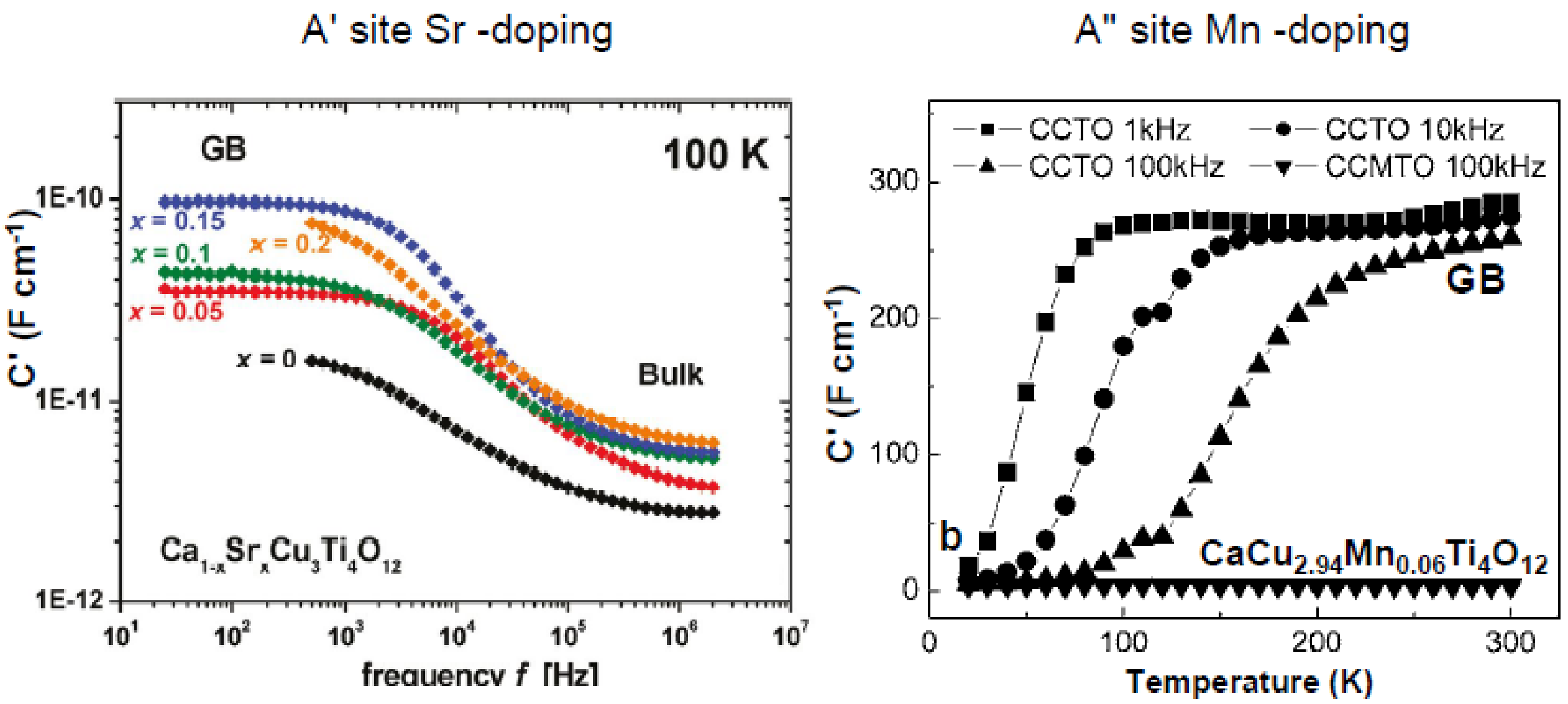

**Figure 7.** Effects of A'-site Sr- and A''-site Mn-doping on the dielectric properties of CCTO. GB and bulk (b) capacitance plateaus are indicated. **Left panel:** *C*' vs *f* plots for $Ca_{1-x}Sr_xCu_3Ti_4O_{12}$ ceramics with various *x* doping level. **Right panel:** *C*' vs *T* plots for $CaCu_{2.94}Mn_{0.06}Ti_4O_{12}$. Images reproduced from [5] with permission from the American Chemical Society (ACS) and from [28] with permission from the American Institute of Physics (AIP).

In addition to the two examples shown in Figure 7, further articles in the literature report on an increase in the intrinsic bulk dielectric permittivity in $Na_{1/2}Bi_{1/2}Cu_3Ti_4O_{12}$ [35], an increase in GB permittivity in $CaCu_3Ti_{4-x}Nb_xO_{12+x/2}$ [36], and a decrease in GB permittivity in $La_{2/3}Cu_3Ti_4O_{12}$, $Y_{2/3}Cu_3Ti_4O_{12}$ and $Bi_{2/3}Cu_3Ti_4O_{12}$ [37]. In a recent and very comprehensive study it has been shown that the full rare-earth (RE) series exists [$(RE)_{2/3}Cu_3Ti_4O_{12}$], where the ceramics of this series all show an IBLC structure with the Pr containing ceramic showing the highest GB capacitance [38]. There are further examples in the literature reporting on a wealth of different doping mechanisms such as A'-site La-doping [39], A''-site Mg-doping [40], and B-site Ta, Zr, Cr, Al- or Mn-doping [41-46]. The summary in Figure 8 for several doping cations shows the distinct influence of the dopant on the IBLC structure in terms of variations in $R_{GB}$ and $R_b$ [26].

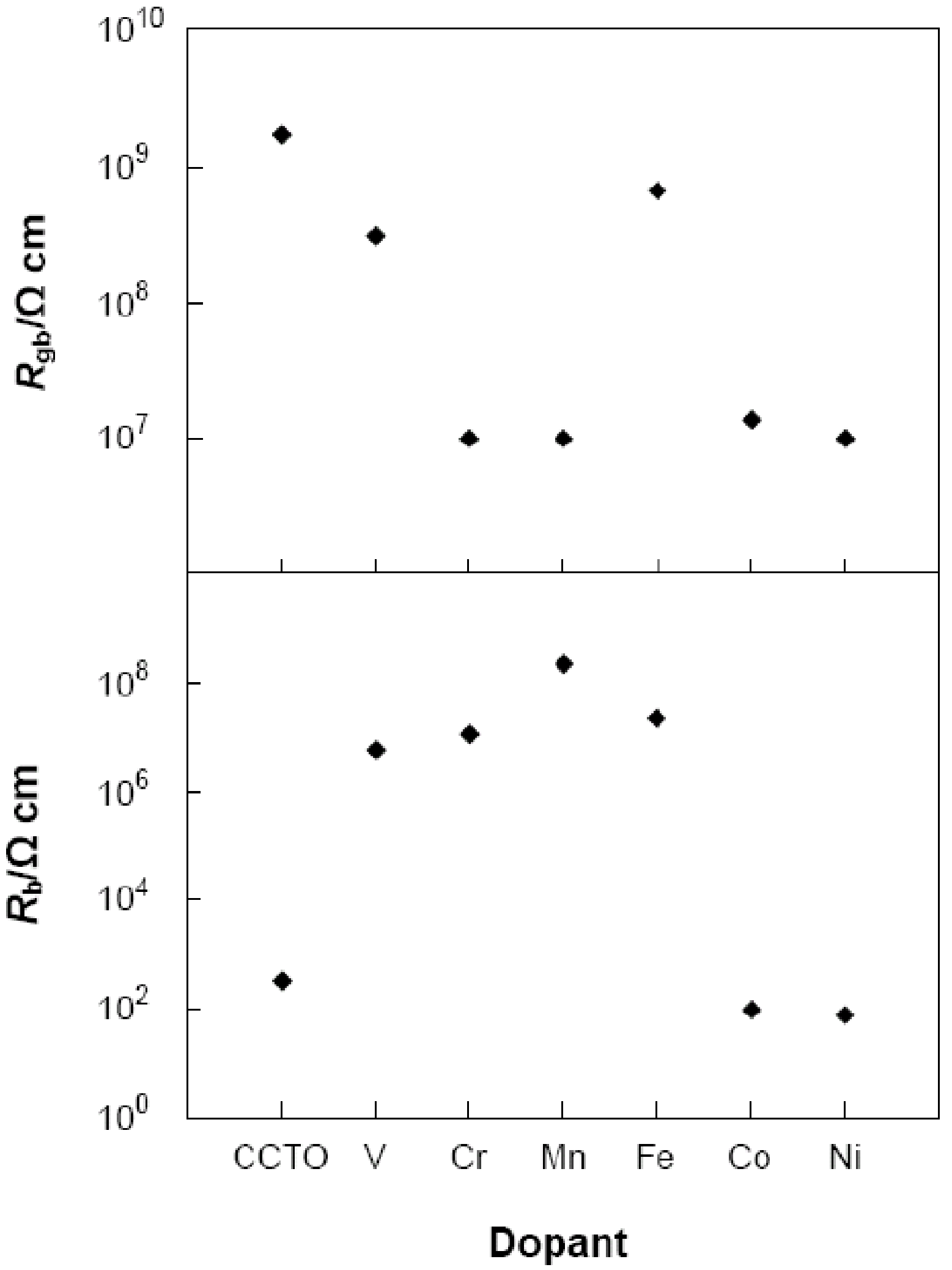

**Figure 8.** Comparison of $R_{GB}$ and $R_b$ at room temperature for undoped CCTO and $x = 0.08$ doped compositions, with the doping element shown. Images reproduced from [47], with permission from M.C. Ferrarelli.

# III. Defect Chemistry within the CaO-CuO-TiO$_2$ Phase Triangle

## A. The CaO-CuO-TiO$_2$ Ternary Phase Diagram

Powders of various compositions within the ternary CaO-CuO-TiO$_2$ phase diagram (Figure 9) were synthesised by repeated grinding and heating at 1000 °C from different amounts of dried precursor oxides of CaCO$_3$, CuO and TiO$_2$. All such compositions are indicated by (black) dots in Figure 9. A small portion of each powder was heated at 1100 °C for 12 h to enable investigation of the phase diagram for materials fired at both 1000 and 1100 °C. The latter is the most frequently used ceramic densification-sintering temperature, whereas CCTO powder synthesis is most commonly achieved at ≈ 1000 °C.

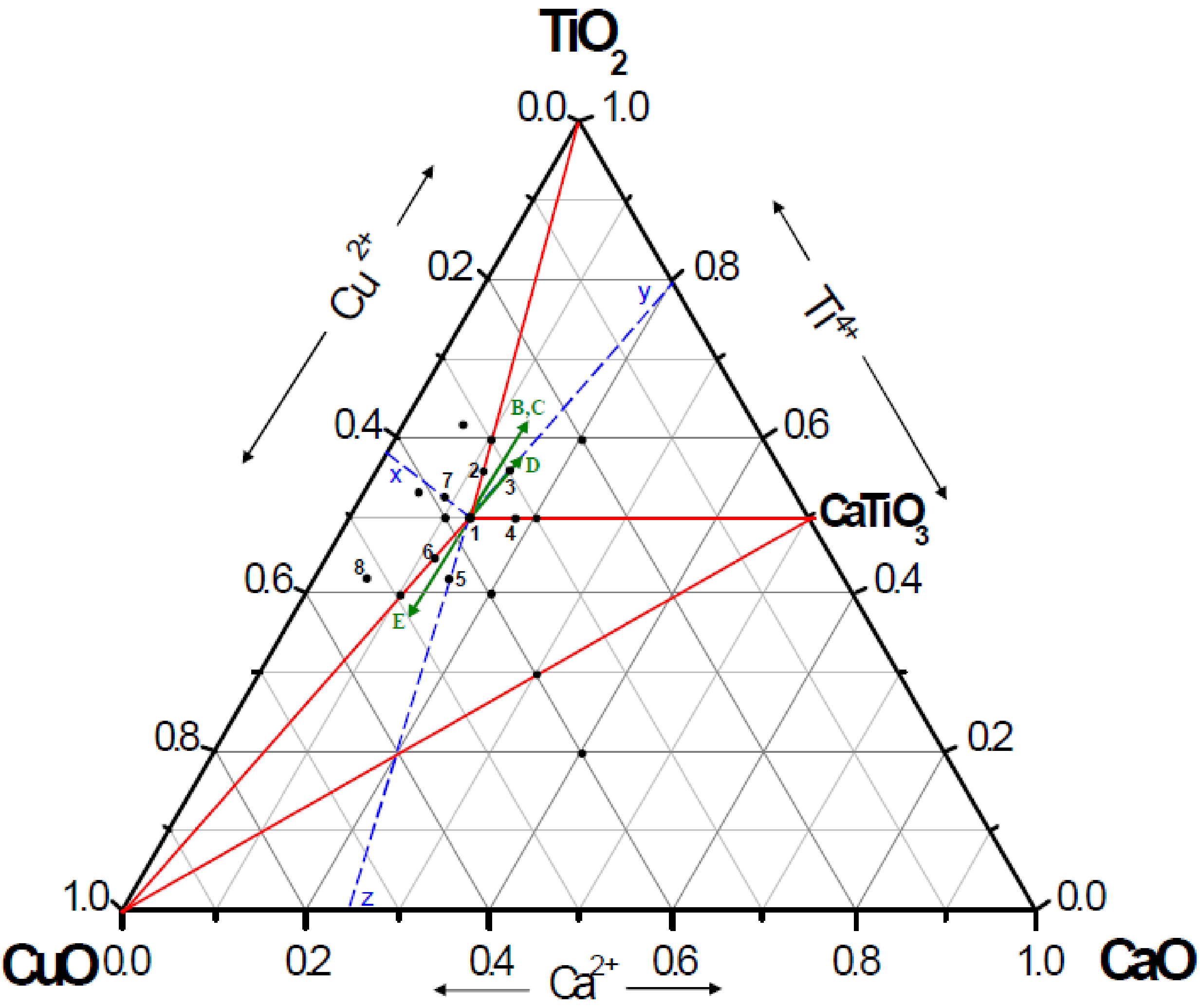

**Figure 9.** Ternary CaO-CuO-TiO$_2$ phase diagram. Black dots represent the compositions investigated. Numbers represent compositions analyzed in more detail by XRD and IS. Dashed blue lines represent CCTO compositions with potential Ca, Cu or Ti loss: $Ca_{1-x}Cu_3Ti_4O_{12}$, $CaCu_{3-y}Ti_4O_{12}$, $CaCu_3Ti_{4-z}O_{12}$, respectively. Green arrows indicate the directions of the potential solid solutions mentioned above in section II./C.

All labelled (numbered) compositions in Figure 9 were investigated in more detail by XRD lattice parameter determination and IS was carried out on pressed pellets sintered at 1100 °C. For compositions 1 – 8 the powder synthesis and pellet sintering processes were each carried out simultaneously for all compositions to exclude differences in the processing conditions.

Compositions 1 - 8 in Figure 9 were chosen in a way such that the volume concentration of potential secondary phases like CuO, $TiO_2$ and $CaTiO_3$ would always be well below the percolation threshold for secondary phases randomly distributed within the primary CCTO matrix. This would ensure the dielectric response in all labelled compositions to always be dominated by the CCTO phase. The theoretical percolation threshold was calculated to be about 16% [48], whereas experimental values are commonly reported to be up to ≈ 20% [49].

The only single-phase composition detected within the main triangle was the CCTO phase, marked as composition 1. $CaTiO_3$ (CTO) was the only significant single-phase on the CaO-$TiO_2$ binary joint. Along all the solid (red) lines within the main triangle two phases are present, which are the two located on both ends of each line: e.g. CCTO and $TiO_2$ are present on the line connecting CCTO (composition 1) and the $TiO_2$ corner.

The various secondary triangular regions within the main triangle all contain three phases, which are the ones that span out the secondary triangles: e.g. CCTO, CTO and $TiO_2$ are present inside the triangle with corners CCTO, CTO and $TiO_2$. No direct evidence for two CCTO compositions was found, i.e. for GB and bulk CCTO phases, in agreement with recent work [50].

The phase relationships detected are consistent with no obvious solid solution being evident and CCTO and CTO are the only non-trivial single-phase compositions. Potential solid solutions causing non-stoichiometry must therefore be small. The XRD patterns for all numbered compositions in Figure 9, heat treated at 1100 °C, are displayed in Figure 10. The phase diagrams for heat treatments at 1000 and 1100 °C were found to be identical with one exception: no crystalline CuO phase could be detected at 1100 °C in any composition by powder XRD, whereas at 1000 °C crystalline CuO was detected in all compositions where it was expected according to the phase diagram. It is suggested that CuO may partially melt at 1100 °C, which was confirmed on CuO-rich pellets where a Cu-rich phase precipitated out of the pellets during sintering at 1100 °C and remained on the Pt foil after cooling. The absence of CuO phases in these pellets is clearly demonstrated in the XRD pattern presented in Figure 10.

In section II./.C. several solid solutions constituting potential CCTO defect mechanisms were mentioned to explain differences in GB and bulk regions, where all such mechanisms are indicated by (green) arrows in Figure 9. For demonstration purposes the length of the (green) arrows is depicted as being much higher than the expected small extent to which the solid solutions would be developed.

Problems were encountered with the triangle having corners CuO, CTO and CaO, where the phase relationships could not be determined unambiguously. The compositions in this triangle were subjected to partial melting of CuO and possibly Ca containing phases other than CaO were formed. The single-phase compound $Ca_2CuO$ on the CaO-CuO binary joint was postulated to exist in a recent work on powders heated at 950 °C [50] but could not be confirmed here for powders prepared at 1000 and 1100 °C.

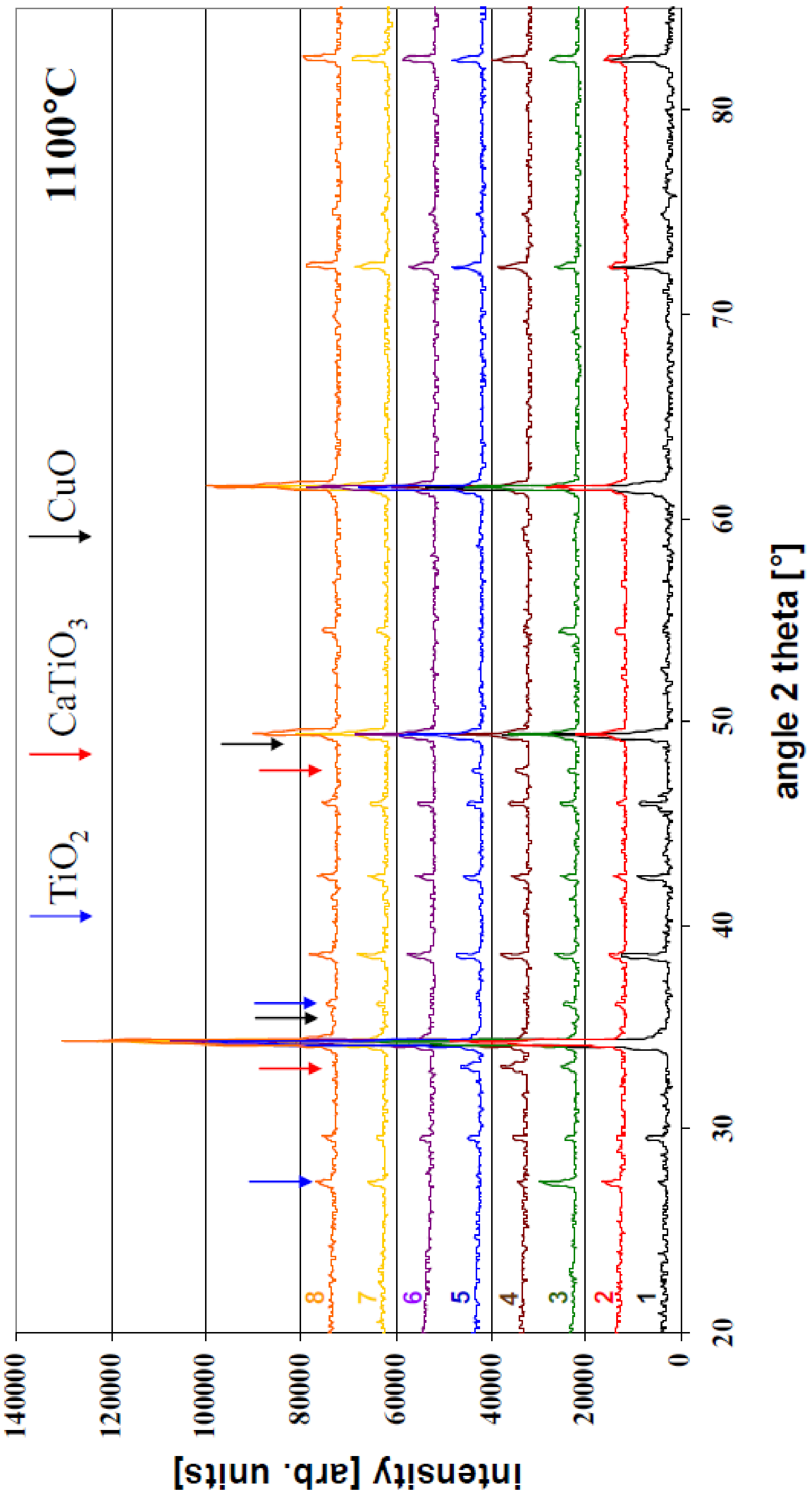

**Figure 10.** Powder XRD pattern of compositions 1-8 fired at 1100°C. Blue, red and black arrows correspond to the most intence $TiO_2$, $CaTiO_3$ and CuO reflectios respectively.

## B. Impedance Spectroscopy on Various CaO-CuO-$TiO_2$ Compositions

Pellets pressed from powders of compositions 1 – 8 were sintered at 1100 °C, polished and covered on both sides with Au electrodes using dc sputtering for IS measurements between 10 - 500 K at 1 Hz – 2 MHz. Figure 11 shows the data collected at 80 K in the format of the real part of dielectric permittivity $\varepsilon'$ vs frequency $f$ (Figure 11a) and imaginary part of the dielectric modulus $M''$ vs $f$ (Figure 11b).

The impedance data for compositions 1 - 5 were fitted to the equivalent circuit model shown in Figure 11a, which had been successfully applied before for CCTO ceramics [18]. The intrinsic bulk dielectric relaxation process was fitted with a non-ideal R-CPE-C circuit (R1-CPE1-C1). The use of this circuit has been shown previously to be appropriate to account for non-ideal bulk relaxations in electroceramics [51]. The GB relaxation was fitted with a well-established non-ideal R-CPE element (R2-CPE2) and the data and model in Figure 11a and b show good agreement. A successful equivalent circuit fit was also obtained for composition 6; however, the fitting errors were larger, which may be due to less ideal dielectric behaviour. Figure 11a displays low and high-$f$ plateaus as a result of electronic heterogeneity, which had been mentioned above to represent GB ($\varepsilon_{GB}$) and bulk ($\varepsilon_b$) permittivity respectively. It is obvious that different secondary phases have a distinct influence on the dielectric behaviour of CCTO and the IS data fall into two categories: Compositions 7 and 8 vary from 1 – 5, whereas 6 shows intermediate dielectric response and may fall in between.

a)

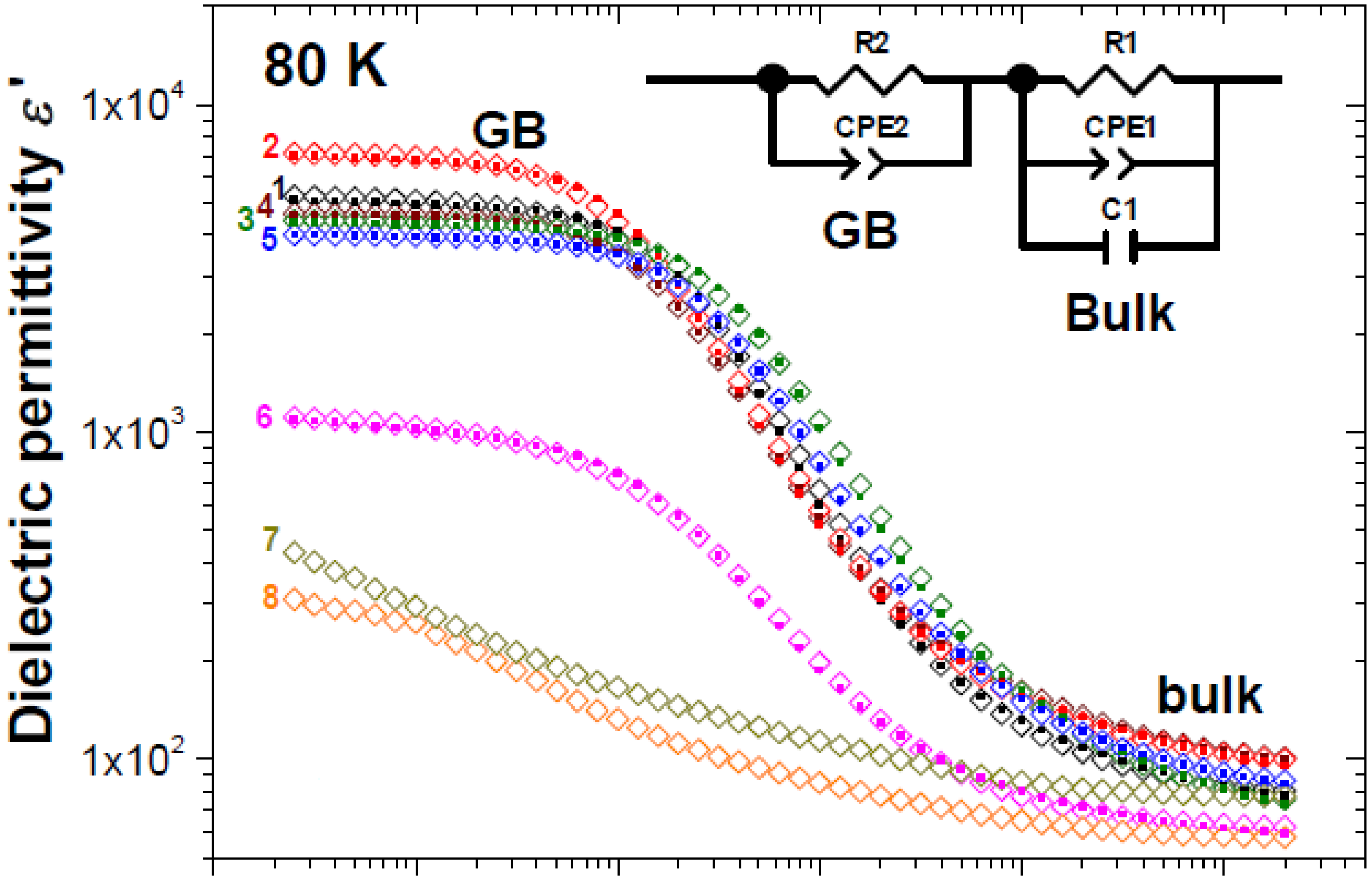

*Figure 11 to be continued on next page.*

b)

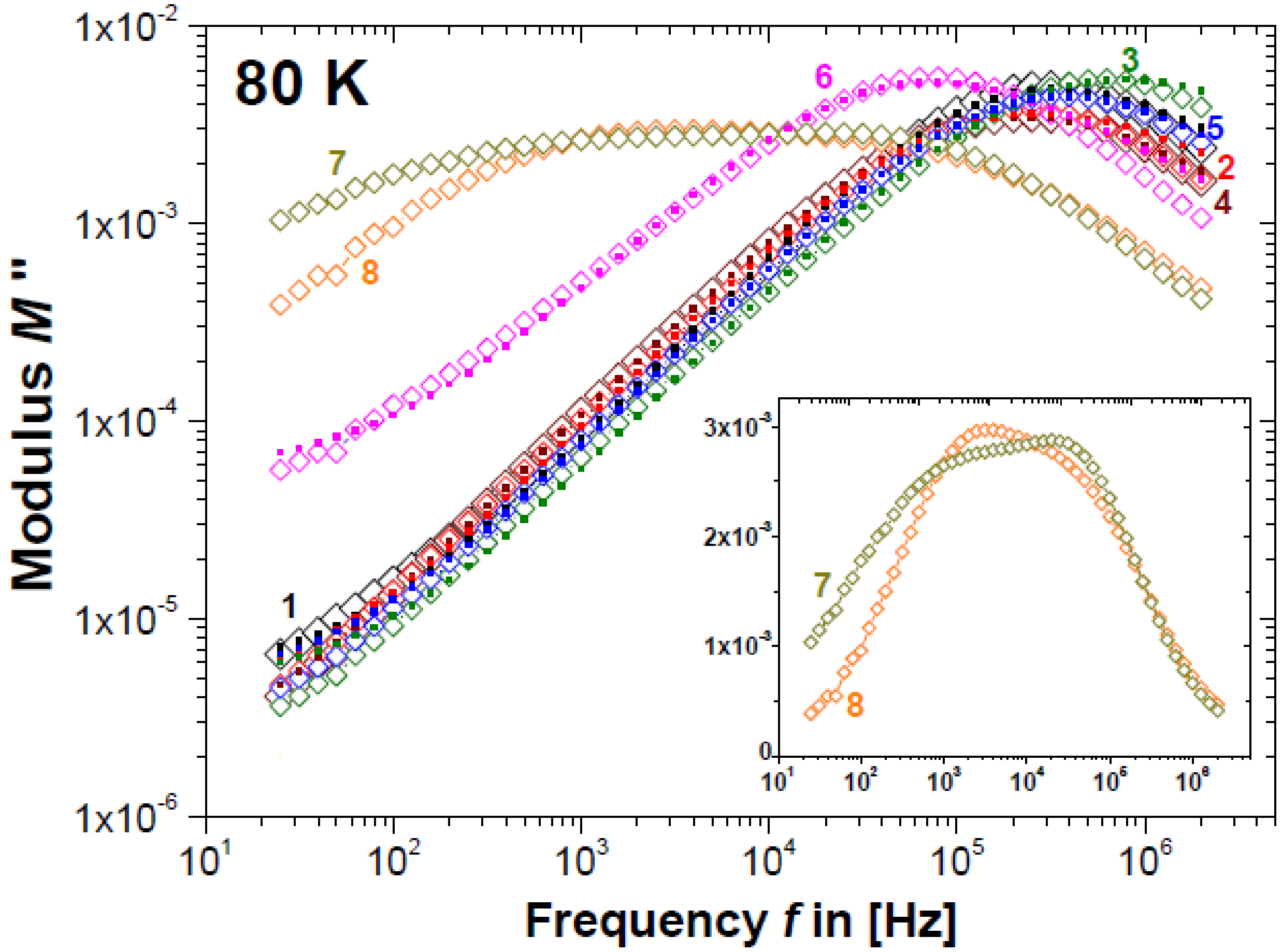

**Figure 11.** IS data for compositions 1 - 8 at 80 K **(a)** *ε*' vs *f*, and **(b)** *M*" vs *f*. Open diamonds (◊) are experimental data, full dots (▪) are fits to the circuit model shown in the inset of Figure 11a.

According to the phase diagram in Figure 9, compositions 7 and 8 belong to the same compositional triangle (CuO-CCTO-$TiO_2$), which would be a plausible explanation for their dielectric distinction. Below in section III./C. it will be shown that this interpretation is incomplete: compositions 5 and 6 also vary from 1 - 4 if their dielectric response is being related to the CCTO lattice parameter ***a***. It appears that it is in fact the presence or absence of the CuO secondary phase that separates the dielectric response into the two categories: i.) 1, 2, 3, 4 (no extra CuO), and ii.) 5, 6, 7 ,8 (extra CuO).

According to Figure 11a the IBLC structure seems to be developed in all compositions, but to different extents. Such variations will be analysed in detail in the following to obtain information on potential defect mechanisms being present. In several previous studies the effects of secondary phases on the CCTO giant dielectric permittivity have already been analysed in detail, although the giant permittivity is extrinsic in nature. This analysis therefore gives phenomenological information only and the investigation of the intrinsic bulk dielectric relaxation is required to gain fundamental insight into the potential intrinsic changes of the CCTO phase induced by the presence of secondary phases. As mentioned above in section II./A. the CCTO intrinsic bulk dielectric relaxation processes may be investigated best by plots of *M*" vs *f*, which are shown in Figure 11b for compositions 1 – 8.

The equivalent two categories are observed as in Figure 11a. For compositions 1 – 5 one high *f* relaxation peak each is displayed, which is indicative of a conventional bulk dielectric relaxation process. Compositions 7 and 8 display a flat or in fact a double peak structure in *M*" vs *f* (see Figure 11b inset) and the bulk dielectric response seems to be distinctively different, whereas composition 6 may fall in between. Again, this interpretation is incomplete as mentioned above and the categories are i.) 1 - 4 and ii.) 5 - 8. The rather untypical and strongly overlapping "double-peaks" in 7 and 8 both occur at significantly lower *f* when compared to the bulk peaks for 1 – 5, whereas the peak heights (ordinates) are only slightly reduced. Using equation (4) this implies that the two double-peak relaxation processes in 7 and 8 both exhibit significantly increased resistance as compared to the single CCTO bulk relaxation for compositions 1 – 5. The relaxation peak of composition 6 indicates increased resistance as well. From these findings it seems plausible that Cu-rich compositions may display higher resistance and an additional CuO phase drives the bulk CCTO peak to higher *f* towards the GB relaxation. This may serve as an explanation for the formation of the IBLC structure: in a recent study it has been shown that Cu segregates out of the CCTO grains during the sintering process [52], which can lead to Cu-rich resistive GBs and Cu-deficient conducting bulk regions.

Since the typical high *f* bulk dielectric relaxation peak displayed in 1 – 5 is missing in 7 and 8 (Figure 11b), it may be argued that the peak at higher *f* of the two double-peaks is more likely to represent the intrinsic bulk CCTO relaxation of composition 7 and 8. The origin of the second peak at slightly lower *f* is unclear and may represent a modified GB contribution or a secondary phase. A complete absence of the CCTO bulk relaxation is unrealistic due to the XRD patterns clearly displaying the CCTO phase, which in fact strongly dominates over all secondary phases (see Figure 10). Due to the double- peak overlap and considerable peak broadening the data for 7 and 8 cannot be fitted by an equivalent circuit model based on RC elements as for 1 – 6.

## C. Combined Analysis of Lattice and Dielectric Parameters

Summarising results from the above section, indications have been found in the impedance spectra that potential non-stoichiometry in CCTO may involve Cu. In this section clear trends of bulk resistance $R_b$ vs lattice parameter ***a*** and bulk dielectric permittivity $\varepsilon_b$ vs ***a*** are presented that reveal the existence of at least two different defect mechanisms and associated solid solution in CCTO.

### *1. Solid Solution in Cu-Deficient Compositions*

Figure 12a shows a plot of bulk resistance $R_b$ obtained from equivalent circuit fits vs the lattice parameter ***a***, for compositions 1 - 6. ***a*** was determined from high resolution powder XRD pattern, where all diffraction peak positions were normalized to a Si standard and at least 4 peaks at high diffraction angles were used in each case to obtain ***a*** from a fitting routine using commercial WinXPOW® software.

Dielectric data were collected from pellets sintered at 1100 ºC. For compositions 1 – 4, $R_b$ at 150 K shows a modest and approximately linear increase with ***a***. It is unclear at this point whether this increase of $R_b$ with ***a*** represents a solid solution because the increase is small and may simply be a consequence of the increasing atomic distance of the cations. With increasing ***a*** the orbital overlap may reduce leading to higher $R_b$.

Compositions 5 and 6 do not follow the same trend and their $R_b$ values seem to be elevated. The curves of $R_b$ vs ***a*** for 1 – 6 at alternative temperatures of 20, 50 and 100 K all show equivalent features. According to the phase diagram in Figure 9, compositions 5 and 6 are expected to contain secondary CuO phase, whereas compositions 2 – 4 (and possibly 1) are expected to be Cu-deficient. The presence of extra Cu in 5 and 6 seems to lead to higher bulk CCTO resistance above the trend observed for 1 - 4. This confirms the notion that highly resistive GB regions may contain higher Cu content.

a)

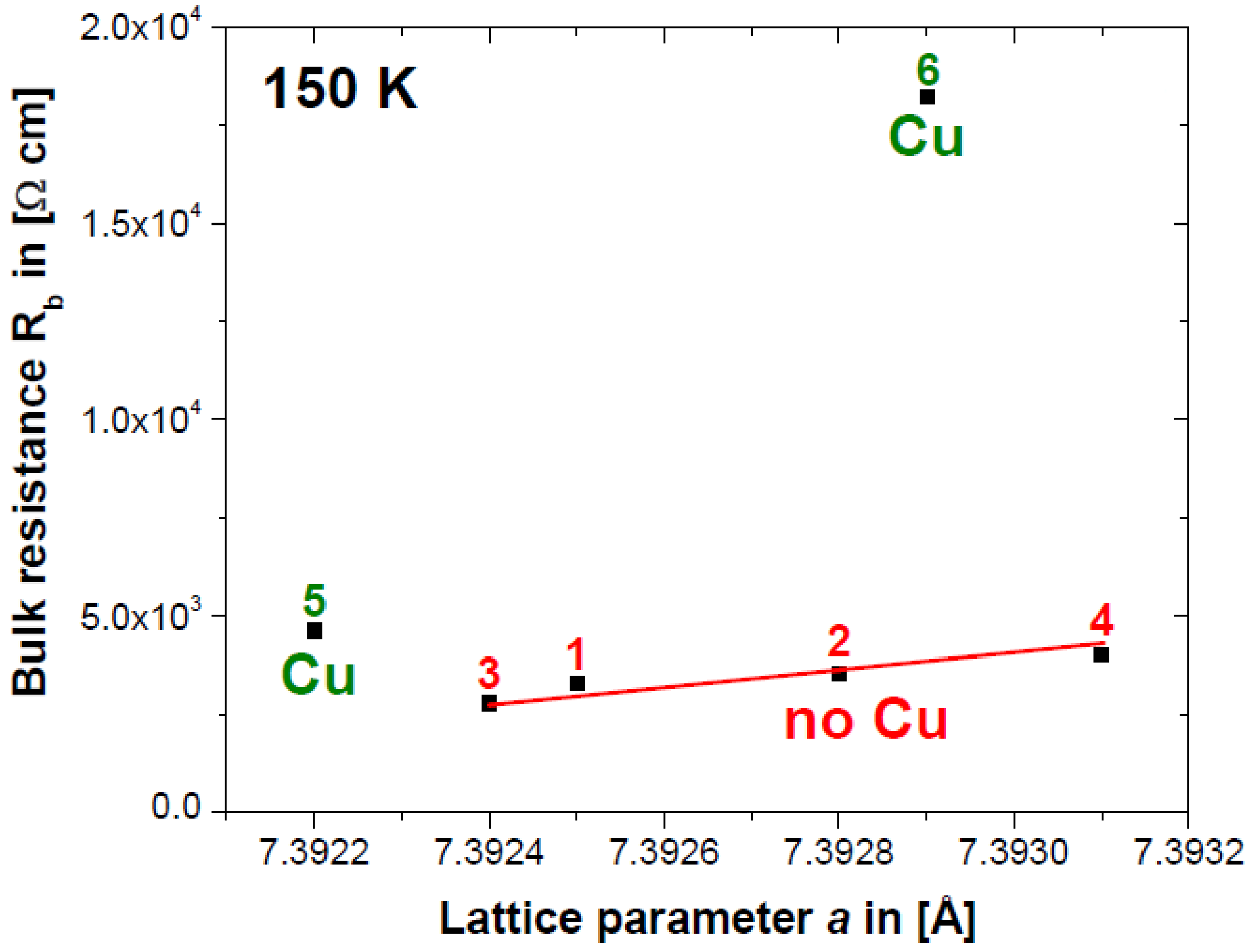

b)

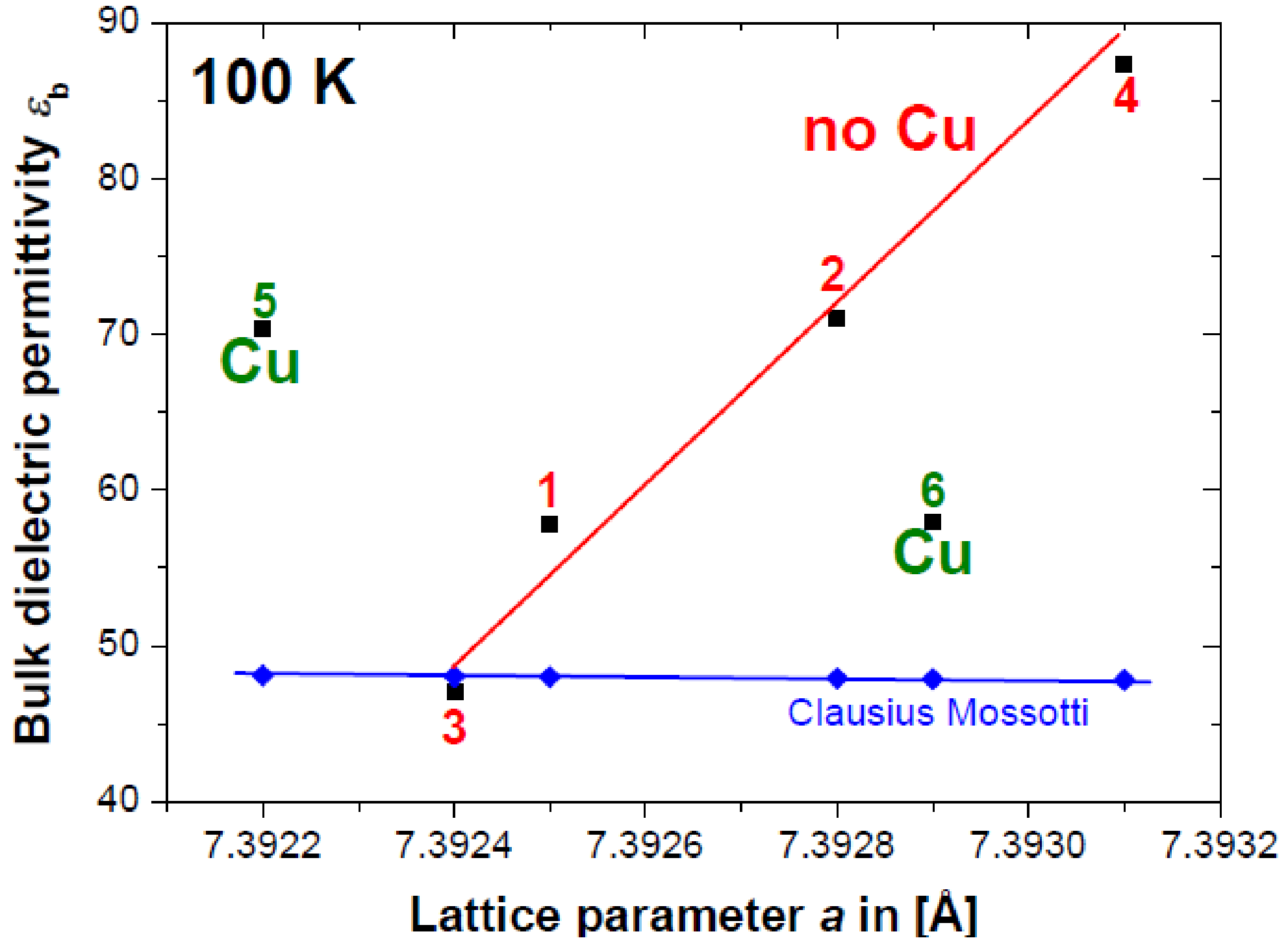

**Figure 12.** Trends for compositions 1 – 6 of **(a)** bulk resistance $R_b$ vs lattice parameter *a* (■), and **(b)** bulk dielectric permittivity $\varepsilon_b$ vs *a* (■). Estimations for $\varepsilon_b$ in CCTO (♦) were obtained from the Clausius Mossotti equation. $R_b$ and $\varepsilon_b$ were obtained from equivalent circuit fits.

The trend of dielectric permittivity $\varepsilon_b$ obtained from equivalent circuit fits vs ***a*** is presented in Figure 12b, where dielectric data had been collected at 100 K. The main feature of this data is the increase of dielectric permittivity $\varepsilon_b$ with ***a*** for compositions 1 – 4, which cannot be explained by simple lattice expansion. The Clausius Mossotti equation predicts approximately constant $\varepsilon_b \approx 48$ considering the small changes in ***a*** observed and taking into account the atom polarizabilities obtained from [53]. The secondary phases in 1 – 4, $CaTiO_3$ and/or $TiO_2$, both have a lower permittivity than CCTO and cannot cause such an anomalous increase in $\varepsilon_b$ directly, but may well influence the defect mechanism in the CCTO bulk phase responsible for the $\varepsilon_b$ increase of up to ≈ 100 above Clausius – Mossotti predictions.

In the case that secondary phase CuO would simply extend the existing defect mechanism for 1 – 4 to higher Cu contents, compositions 5 and 6 would be expected to fall onto the same line as 1 – 4, which is clearly not the case, and one or more additional defect mechanisms may well exist. The mechanism detected for 1 – 4 may not involve large differences in Cu content, because in this case a larger variation of $R_b$ with ***a*** would be likely than that detected in Figure 12a.

### *2. Solid Solution in Cu-Rich Compositions*

To elucidate the role of secondary CuO phase on the CCTO bulk resistance, the $R_b$ values for all compositions containing extra CuO (5 – 8) were estimated from the relaxation peak height and peak position in the $M''$ vs $f$ plots (Figure 11) by using equations (4) and (5). For the double-peaks in compositions 7 and 8 estimates were taken from the peak at higher $f$ in the $M''$ vs $f$ plots, which was argued above to be more likely to represent the CCTO bulk relaxation. The estimated $R_b$ values obtained from $M''$ vs $f$ plots for 5 and 6 were compared to the respective $R_b$ values from the equivalent circuit fits (section III./C./1.).

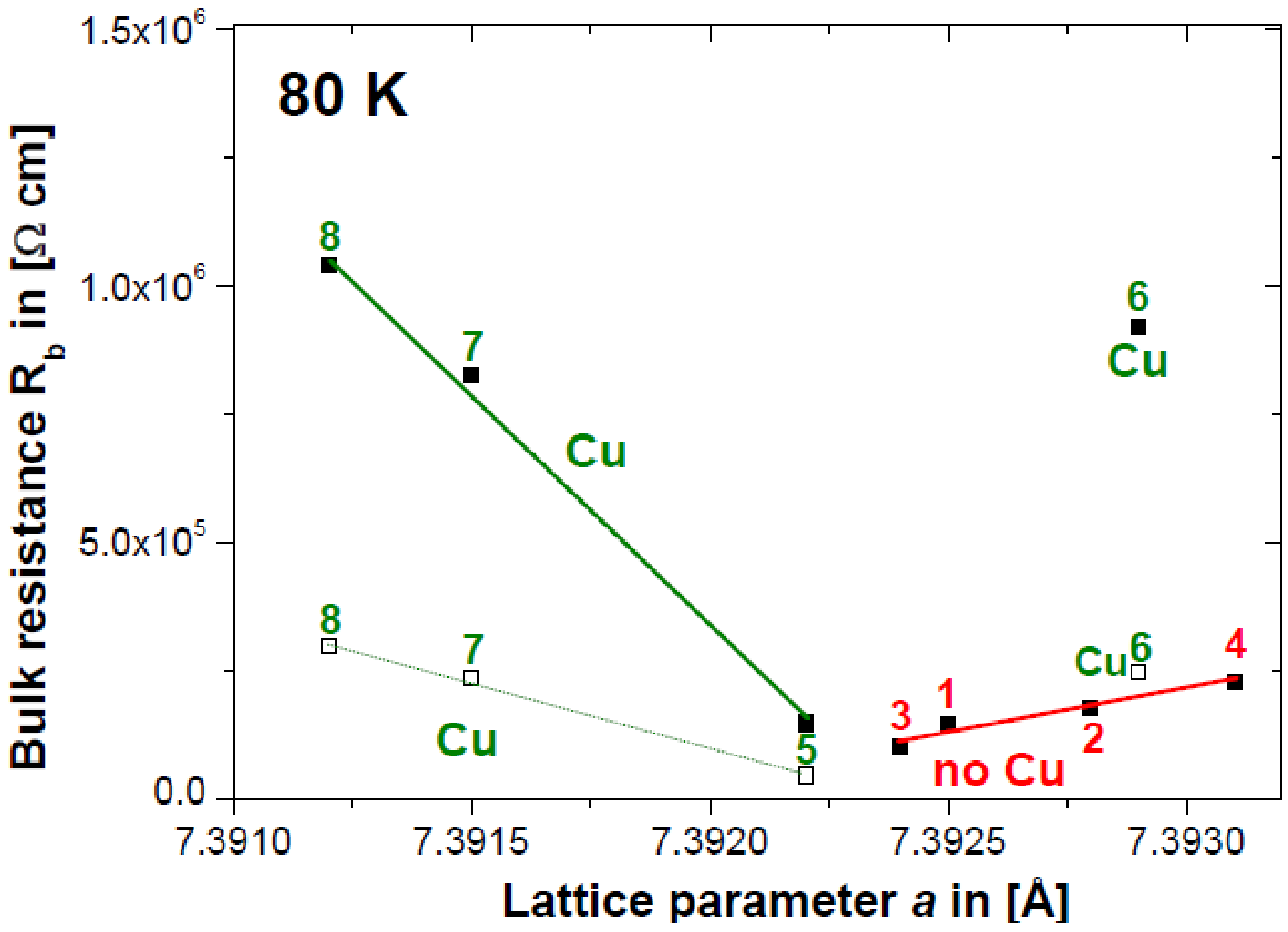

**Figure 13.** Trends of $R_b$ vs $a$ for all compositions (1 – 8) from equivalent circuit fits (■) and from estimates obtained from $M''$ vs $f$ relaxation peak analysis (□). For compositions 7 and 8 more realistic $R_b$ values (■) were estimated.

It was found that at 80 K the estimated $R_b$ values were smaller by a factor of 3.23 (composition 5) and 3.72 (composition 6), which can be explained by the non-ideality of the relaxation processes leading to a broadening of the relaxation peaks in *M*'' vs *f*. This, in turn, leads to a reduction in peak height and an overestimate of $\varepsilon_b$ according to equation (5). The error made by the $R_b$ estimates from *M*'' vs *f* plots was approximated to be a uniform factor of ≈ 3.5. Thus, the estimated values (□ in Figure 13) for 7 and 8 were multiplied by a factor of 3.5 to account for the estimation error.

Both, the uncorrected and corrected $R_b$ values for 7 and 8 at 80 K plotted vs ***a*** (■ in Figure 13) approximately fall onto one line with composition 5 and a significant increase of $\rho$ with decreasing ***a*** is found. Composition 6 is an exception to this trend with the reasons being unclear. Still, the significant $R_b$ increase with decreasing ***a*** detected in Cu-rich samples 5, 7 and 8 strongly suggests the existence of a second defect mechanism and an associated solid solution, which may relate the large differences between GB and bulk resistance to the Cu concentration: GB regions would be expected to be Cu-rich and bulk regions Cu-deficient.

## D. Summary and Discussion of the Defect Chemistry in CCTO

Combined analysis of XRD lattice parameter ***a*** and IS data revealed approximately linear relationships between the CCTO bulk resistance $R_b$ vs ***a***, and intrinsic bulk dielectric permittivity $\varepsilon_b$ vs ***a***. For Cu-deficient compositions 1 – 4 an anomaleous increase of the CCTO bulk dielectric permittivity $\varepsilon_b$ with ***a*** was found above the predictions from the Clausius Mossotti equation. A distinct defect mechanism may well exist causing high intrinsic bulk dielectric permittivity.

Compositions 5 – 8 containing a CuO secondary phase did not folllow the same trends. The presence of excess CuO led to increased $R_b$ and indications for a second defect mechanism were found. This second mechanism may explain the large difference between GB and bulk resistance in CCTO in terms of the Cu content. GB regions would be expected to exhibit a higher Cu content as compared to Cu-deficient bulk regions. In compositions 1 – 4 the source of semiconductivity in the respective CCTO bulk phase may well be Cu deficiency too, but it may be independent of the $TiO_2$ and/or $CaTiO_3$ secondary phases since no strong differences in $R_b$ were detected.

The rather small increase of $R_b$ with ***a*** in 1 – 4 (Figure 12a) leads to some further important conclusions: the underlying defect mechanism causing high $\varepsilon_b$ may not involve any large changes in the Cu nor the Ti valencies, in which case larger differences in $R_b$ would be likely. This would exclude major contributions from all mechanisms mentioned in section II./C except mechanism (E), which constitutes a Cu excess compensated by the formation of oxygen vacancies. Since compositions 1 – 4 are all Cu deficient, mechanism (E) is unlikely. Instead, the formation of anti-site defects like Ca-Cu, Cu-Ti or Ca-Ti inter-site cation exchanges may be a more plausible explanation. Evidence for Ca-Cu anti-site disorder has indeed been reported in the literature [54], although it should be noted that here in this work the anti-site defects proposed would serve as an explanation for the increased bulk permittivity $\varepsilon_b$ and not for the extrinsic giant permittivity $\varepsilon_{GB}$. Furthermore, the possibility still exists that at least one of the defect mechanisms A – D may accompany the anti-site defect formation to a small extent.

The trends of $R_b$ vs ***a*** and $\varepsilon_b$ vs ***a*** are relatively uniform for all compositions investigated which may be a result of the identical powder synthesis and pellet sintering conditions employed. The two distinct defect mechanisms detected were only obvious from combined analysis of ***a*** and IS data and cannot be clearly associated to two of the defect mechanisms A - E. From the difficulty of detection it is obvious that the extent to which the defect mechanisms in CCTO ceramics are developed may be small and difficult to control. In the next section it will nevertheless be demonstrated that the extent of both defect mechanisms is directly proportional to the ceramic densification sintering temperature, $T_S$.

# IV. Influence of Sintering Temperature on the IBLC Structure

## A. Dielectric Spectra of CCTO Ceramics Sintered at Various Temperatures

Nominal stoichiometric CCTO powder was synthesized from dried reagents of $CaCO_3$, CuO and $TiO_2$ at 1000 °C. Six pellets were pressed from fresh CCTO powder and each pellet was sintered at a different temperature $T_S$, i.e. $T_S$ = 975, 1000, 1025, 1050, 1075 and 1100 °C on Pt foil for 12 h followed by slow-cooling in the furnace. For dielectric measurements by IS the pellet surfaces were polished and covered with Au electrodes using dc sputtering. The IS data collected at 80 K for all pellets sintered at different $T_S$ are plotted as the real part of the dielectric permittivity $\varepsilon'$ vs frequency $f$ in Figure 14, where two distinct permittivity plateaus are shown representing GB and bulk contributions. It is apparent that the low-$f$ GB plateau value $\varepsilon_{GB}$* increases substantially with increasing $T_S$.

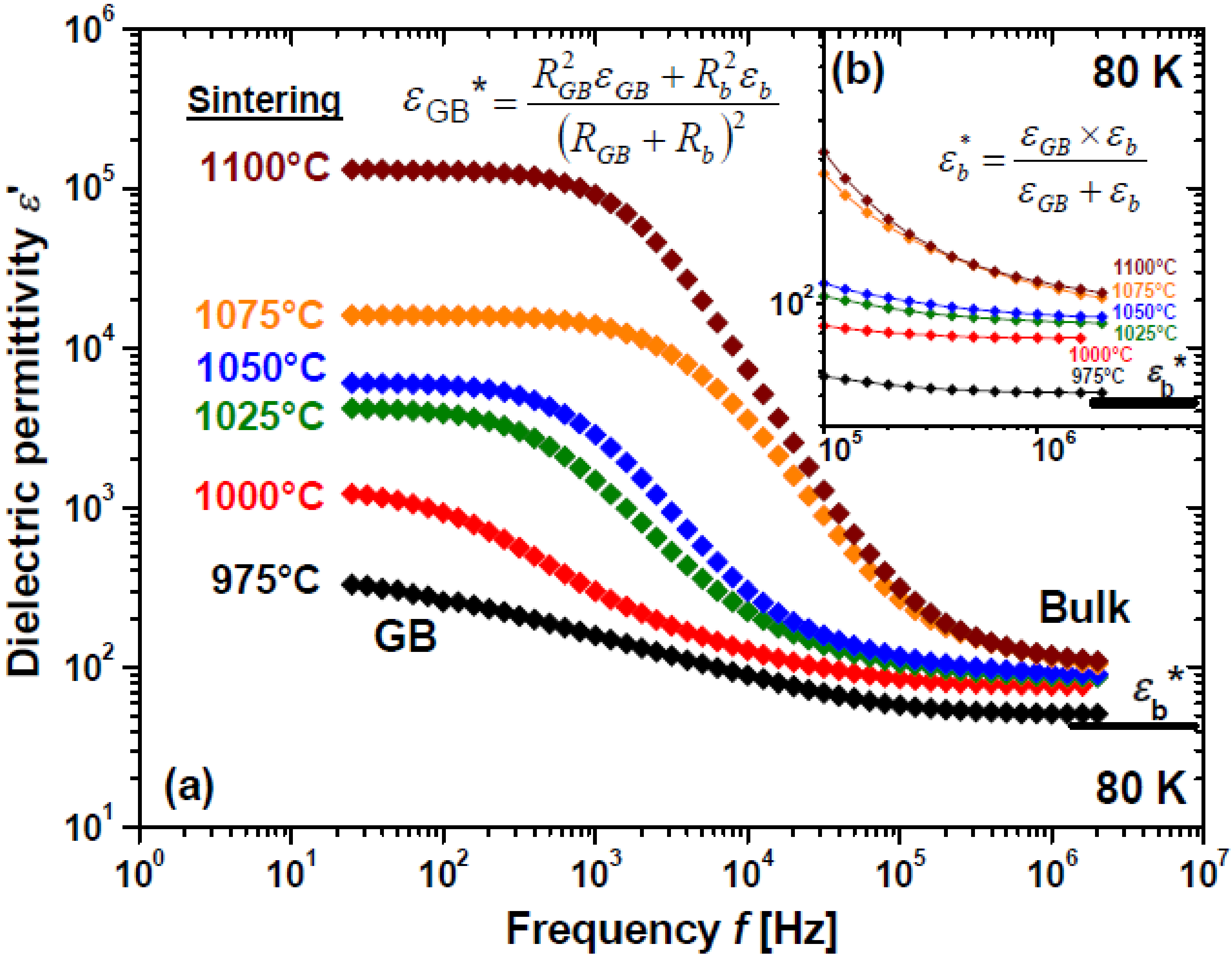

**Figure 14. (a)** $\varepsilon'$ vs $f$ at 80 K for pellets sintered at various $T_S$. The magnitude of the GB plateau $\varepsilon_{GB}$* increases significantly with $T_S$. **(b)** High $f$ behaviour of $\varepsilon'$ vs $f$. Axes are identical to the main figure. The bulk permittivity plateau $\varepsilon_b$* increases with increasing $T_S$. Reproduced from [52] with permission from the European Ceramic Society.

This increase is far too large to be explained by the changes in grain size (as presented in section IV./C.) or by the increase observed in the pellet density from ≈ 80 to ≈ 95 % for $T_S$ = 975 to 1100 °C. Therefore, it can be concluded that the GB permittivity in CCTO shows a substantial increase with increasing $T_S$. A magnification of the Figure 14a data is shown in Figure 14b focusing on high *f*, where the magnitude of the bulk-dominated low permittivity plateau $\varepsilon_b$* increases moderately with $T_S$ as a manifestation of a modest increase of bulk dielectric permittivity $\varepsilon_b$.

As mentioned above in section II./A. the $\varepsilon_{GB}$* and $\varepsilon_b$* plateaus in Figure 14 are composite terms as demonstrated by equations 2 and 3. Such equations are displayed again in Figure 14 now in the permittivity format $\varepsilon'$. Since the $R_{GB}$ and $R_b$ values in the CCTO ceramics investigated here always vary by more than 3 orders of magnitude, $\varepsilon_{GB}$* can be regarded a good estimate for $\varepsilon_{GB}$.

$\varepsilon_b$* is a good estimate for $\varepsilon_b$ only in the case that the two capacitors $C_{GB}$ and $C_b$ are sufficiently different from each other. As can be seen in Figure 14a this was not always the case and the contribution from $\varepsilon_{GB}$ needs to be considered to calculate $\varepsilon_b$ as explained below in section IV./B. The trend of increasing $\varepsilon_{GB}$ with $T_S$ was confirmed from Cole-Cole plots of $\varepsilon''$ vs $\varepsilon'$ in Figure 15 displaying a single arc, where the arc diameter is directly proportional to the magnitude of $\varepsilon_{GB}$. $\varepsilon_{GB}$* corresponds to the *x*-axis ($\varepsilon'$ axis) intercept at low *f*, whereas the high *f* intercept corresponds to $\varepsilon_b$*, which is close to but not a perfect estimate of $\varepsilon_b$.

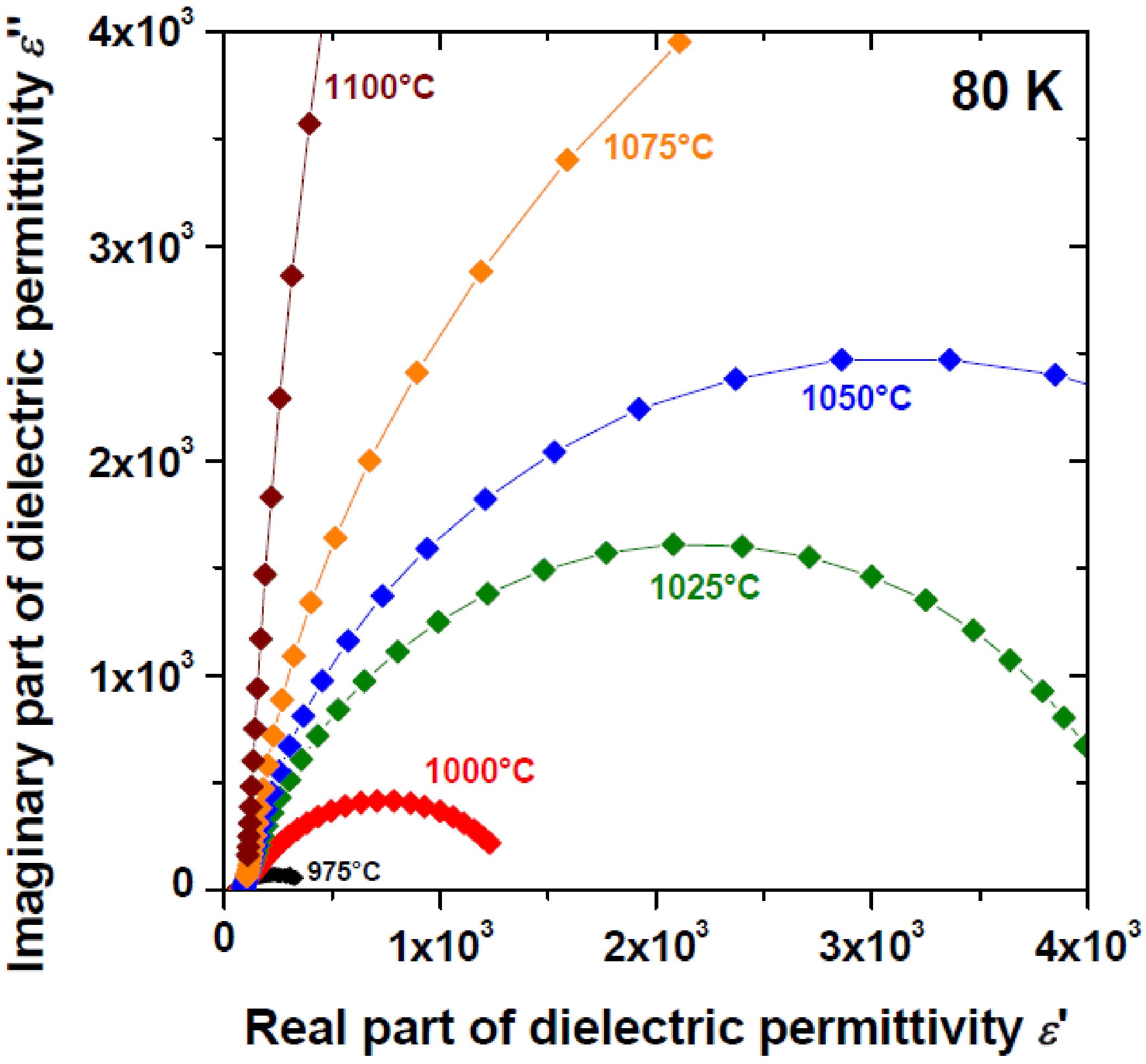

**Figure 15.** Cole-Cole plots of $\varepsilon''$ vs $\varepsilon'$ for CCTO pellets sintered at different temperatures. The diameter of the semicircle is proportional to the GB permittivity. The GB permittivity increases strongly with $T_S$.

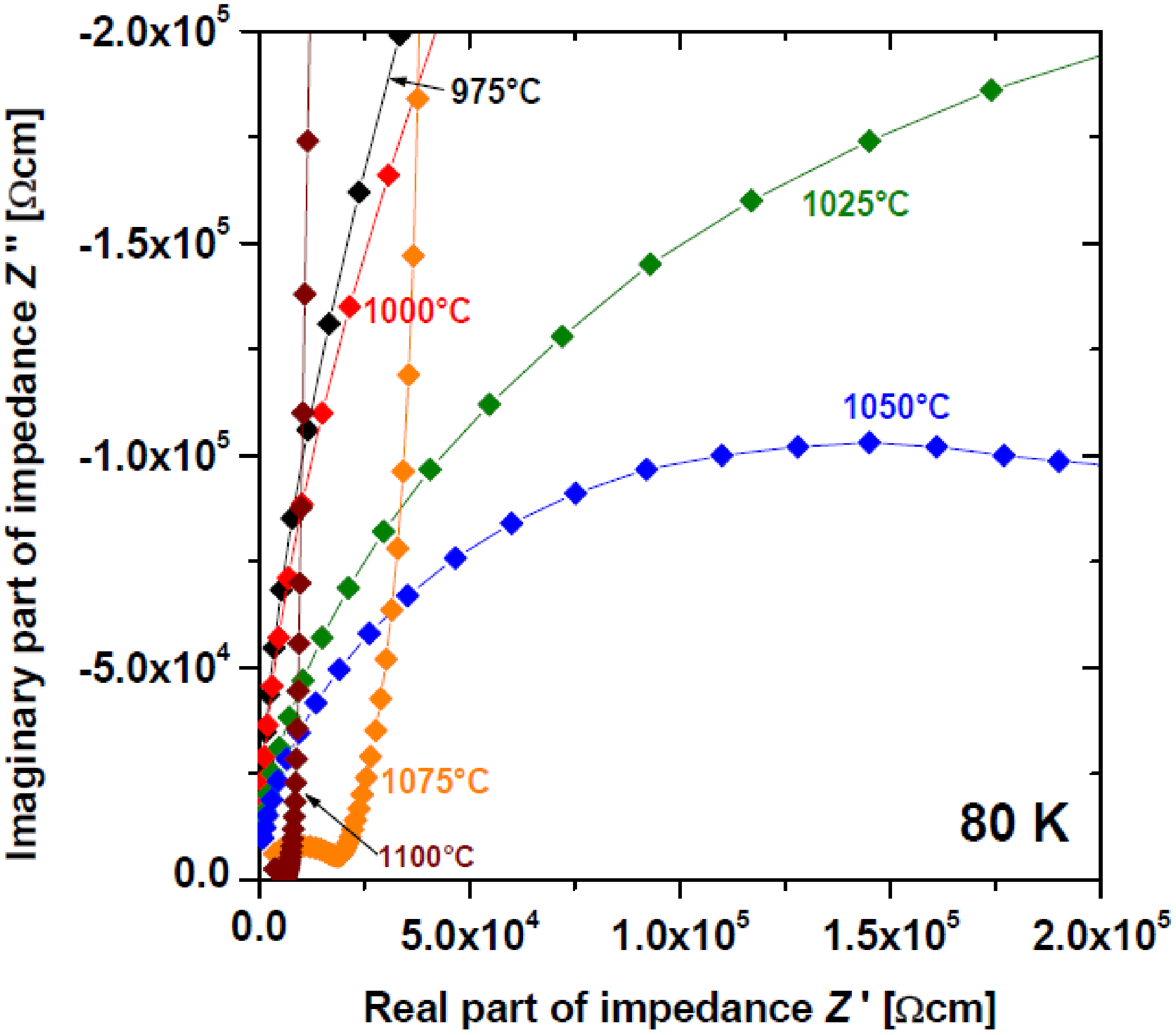

**Figure 16.** Complex-plane plots of *Z*" vs *Z*' for CCTO pellets sintered at different temperature. The diameter of the high *f* semicircle is proportional to the bulk resistance $R_b$. A large decrease of $R_b$ with increasing $T_S$ is observed.

A significant decrease of bulk resistance $R_b$ with $T_S$ is indicated in Figure 16 on complex impedance plots of *Z*" vs *Z*', where the bulk semicircle diameter corresponds to $R_b$. Two semicircles are displayed, which can be seen most clearly for the sample sintered at 1075 °C. At high *f* an almost fully developed bulk semicircle is shown near the origin of the plot, whereas a second, larger GB semicircle is displayed at lower *f*. The GB resistance is too large to be fully resolved on this scale and only the onset of the GB semicircle is shown. In fact, the GB resistance was overlaid by a non-ohmic electrode interface relaxation, which did not allow reliable extraction of GB resistance values.

For the 1100 °C sintered pellet the bulk semicircle near the origin is not fully developed and a non-zero intercept of the data on the real *Z*' axis is observed with only the onset of a small bulk semicircle. At 1075 °C the bulk semicircle is clearly visible, whereas for 1050 and 1025 °C sintered ceramics only a fraction of the bulk semicircle can be displayed and the GB semicircle has disappeared at low *f* due to increased $R_{GB}$. For 1000 and 975 °C sintered pellets only the onset of the bulk semicircle is shown, emphasizing the large increase in $R_b$ with decreasing $T_S$.

## B. Extraction of Dielectric Parameters and Their Trends with Sintering Temperature

All GB and bulk dielectric parameters were determined quantitatively to more clearly assess their trends with $T_S$. Equivalent circuit fitting did not succeed for all samples and, therefore, analysis of the IS data was carried out by the following alternative procedure:

(1) GB permittivity $\varepsilon_{GB}$ was obtained from the high-permittivity plateau $\varepsilon_{GB}$* at low *f* in Figure 14a at 80 K. Contributions from the bulk dielectric relaxation were found to be negligible and therefore no correction was made to the $\varepsilon_{GB}$* plateau value.

(2) Bulk permittivity $\varepsilon_b$ was obtained from the low permittivity plateau $\varepsilon_b$* at high *f* from data taken at 50 K, where the $\varepsilon_1$* bulk plateau is fully developed for all samples. In Figure 14b the samples sintered at 1075 and 1100 $^o$C do not show a full relaxation of $\varepsilon'$ towards the bulk plateau $\varepsilon_b$* at high *f*. $\varepsilon_b$* values were corrected and the precise bulk permittivity $\varepsilon_b$ obtained by taking into account a small contribution from the GB permittivity $\varepsilon_{GB}$, according to equation (3).

(3) Bulk resistance $R_b$ was calculated from the bulk permittivity $\varepsilon_b$ and the frequency of the bulk dielectric relaxation peaks displayed in plots of the electric modulus function *M*'' vs *f* (data not shown here), according to equation (4).

All trends of GB permittivity $\varepsilon_{GB}$, bulk permittivity $\varepsilon_b$ and bulk resistance $R_b$ vs $T_S$ at selected temperatures are shown in Figure 17a-c. The $\varepsilon_{GB}$ values show an exponential increase with $T_S$ by a factor of ≈ 300 (Figure 17a), which is difficult to interpret.

It could reflect intrinsic changes in the GB regions or a geometrical effect associated with GB regions getting thinner with increasing $T_S$. The detected $\varepsilon_{GB}$ increase is in agreement with previous work on CCTO ceramics sintered at different $T_S$ [55]. This increase is significantly more pronounced here as was found in previous studies on 1100 °C sintered CCTO with different sintering time [56, 57]. $T_S$ as opposed to sintering time has a more pronounced influence on the dielectric properties of CCTO, which may reflect the importance of diffusion-related phenomena occurring during the densification sintering process in CCTO ceramics. The intrinsic bulk permittivity $\varepsilon_b$ shows an approximately linear increase with $T_S$ between 1000 - 1100 °C by a factor of ≈ 2 (Figure 17b), which can be associated with intrinsic changes of the bulk phase. Only a small part of this increase can be associated with changes in pellet density.

The rather high intrinsic $\varepsilon_b$ values of ≈ 100 in CCTO and related materials have been associated previously with the formation of an incipient ferroelectric phase [58]. The formation of this phase seems to be particularly pronounced here by increasing $T_S$ from 975 to 1000 °C. As mentioned in section III./C./1., the formation of the incipient ferroelectric phase may possibly be associated with a defect mechanism based on Ca-Cu anti-site defects. Apparently, this mechanism is strongly influenced by $T_S$ leading to a linear increase of $\varepsilon_b$ with increasing $T_S$ and anti-site disorder would gradually evolve with $T_S$ to higher levels.

In Figure 17c the bulk resistance $R_b$ shows an exponential decrease with increasing $T_S$ between 1000 – 1100 °C by a factor of ≈ 1000 – 4000, but no decrease can be observed by increasing $T_S$ from 975 to 1000 °C. Since a clear increase of $\varepsilon_b$ was shown in Figure 17b from 975 to 1000 °C, the $\varepsilon_b$ increase with $T_S$ may rely on a different mechanism in agreement with the arguments proposed in section III./C./2. It is likely that the defect mechanism responsible for the massive reduction of $R_b$ with $T_S$ may involve the Cu concentration as mentioned in section III./C./2. This is confirmed by the notion that increasing Cu segregation out of the CCTO ceramics is the only visible change of the CCTO phase with increasing $T_S$ as will be demonstrated in the next section.

a)

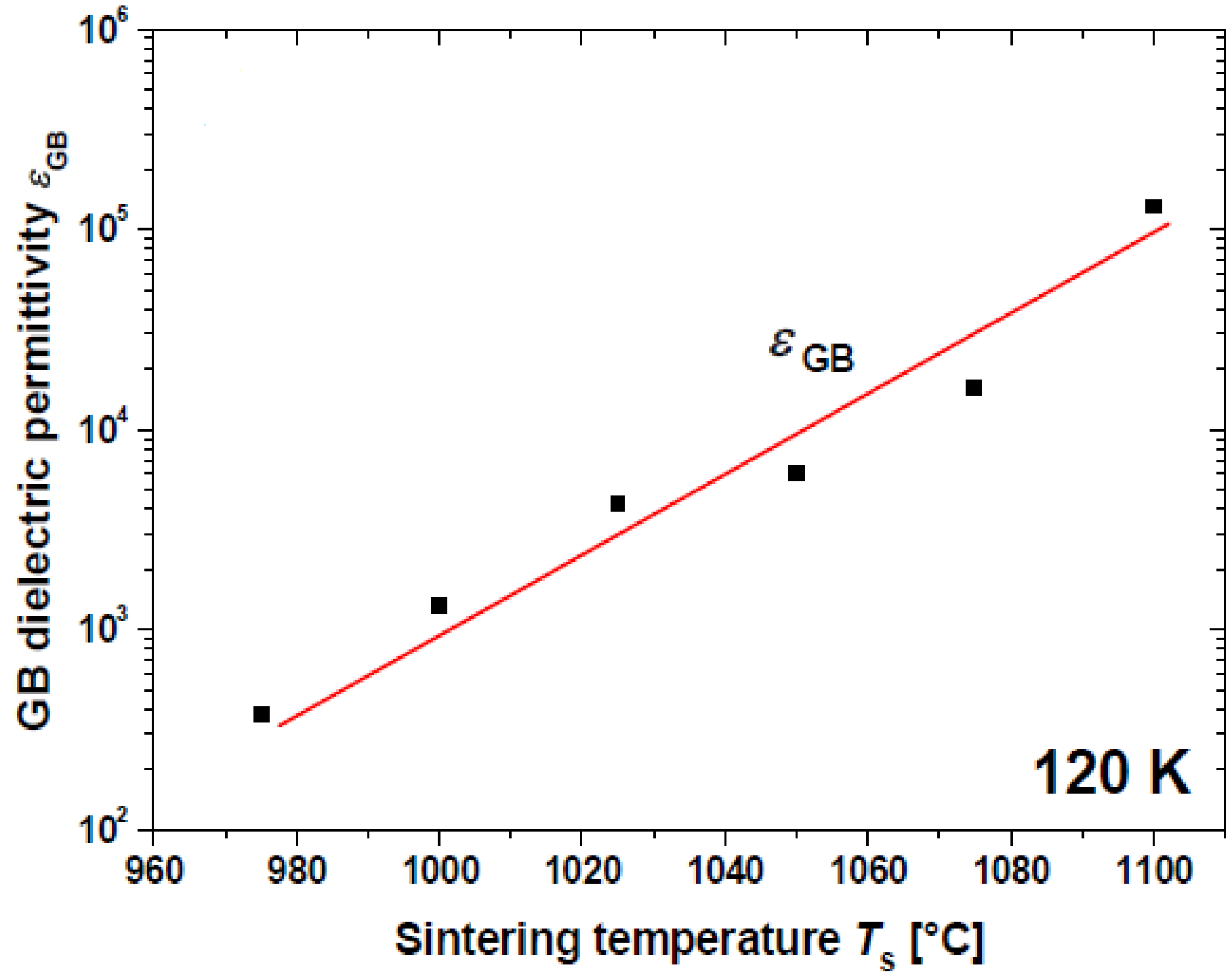

b)

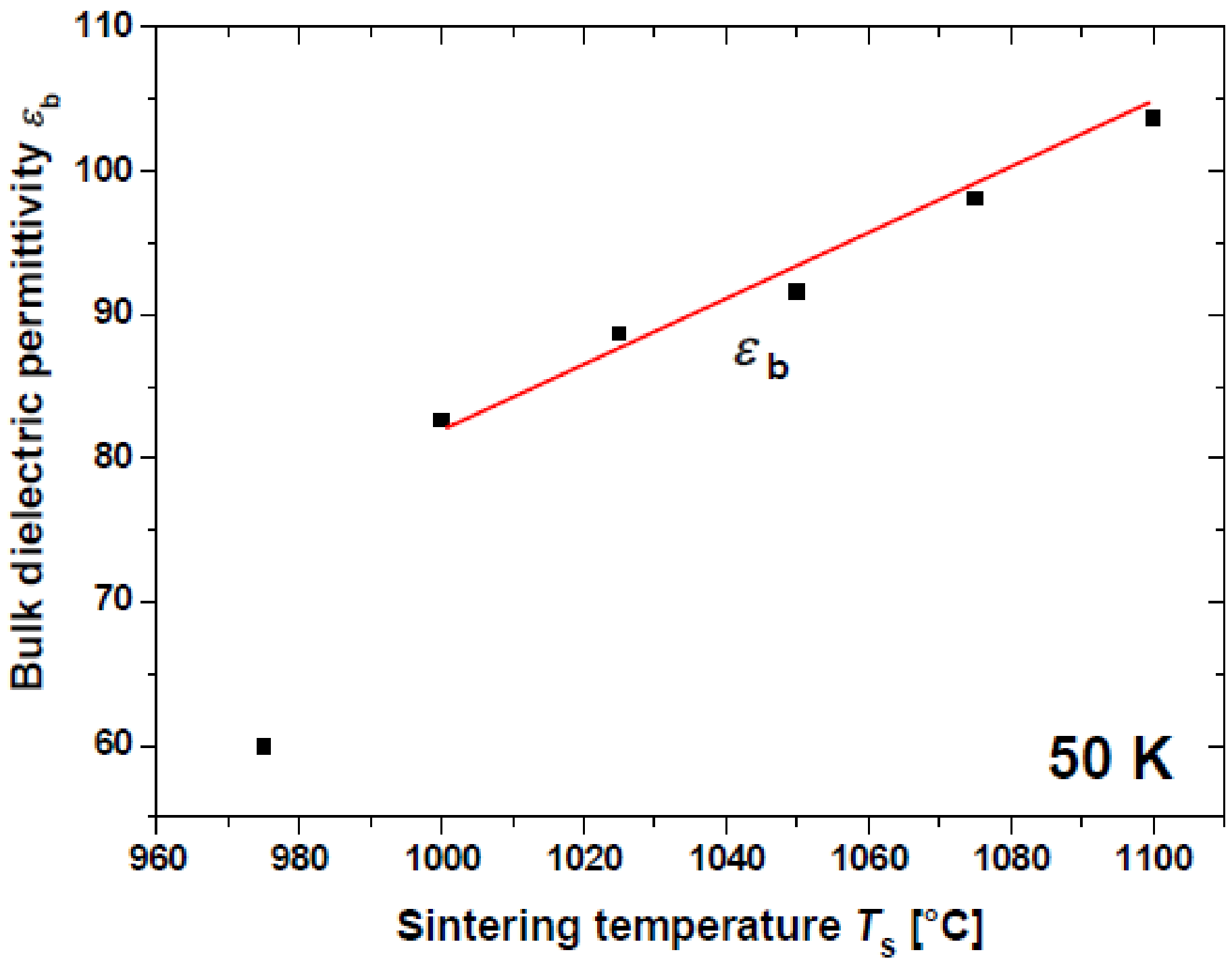

*Figure 17 to be continued on next page.*

c)

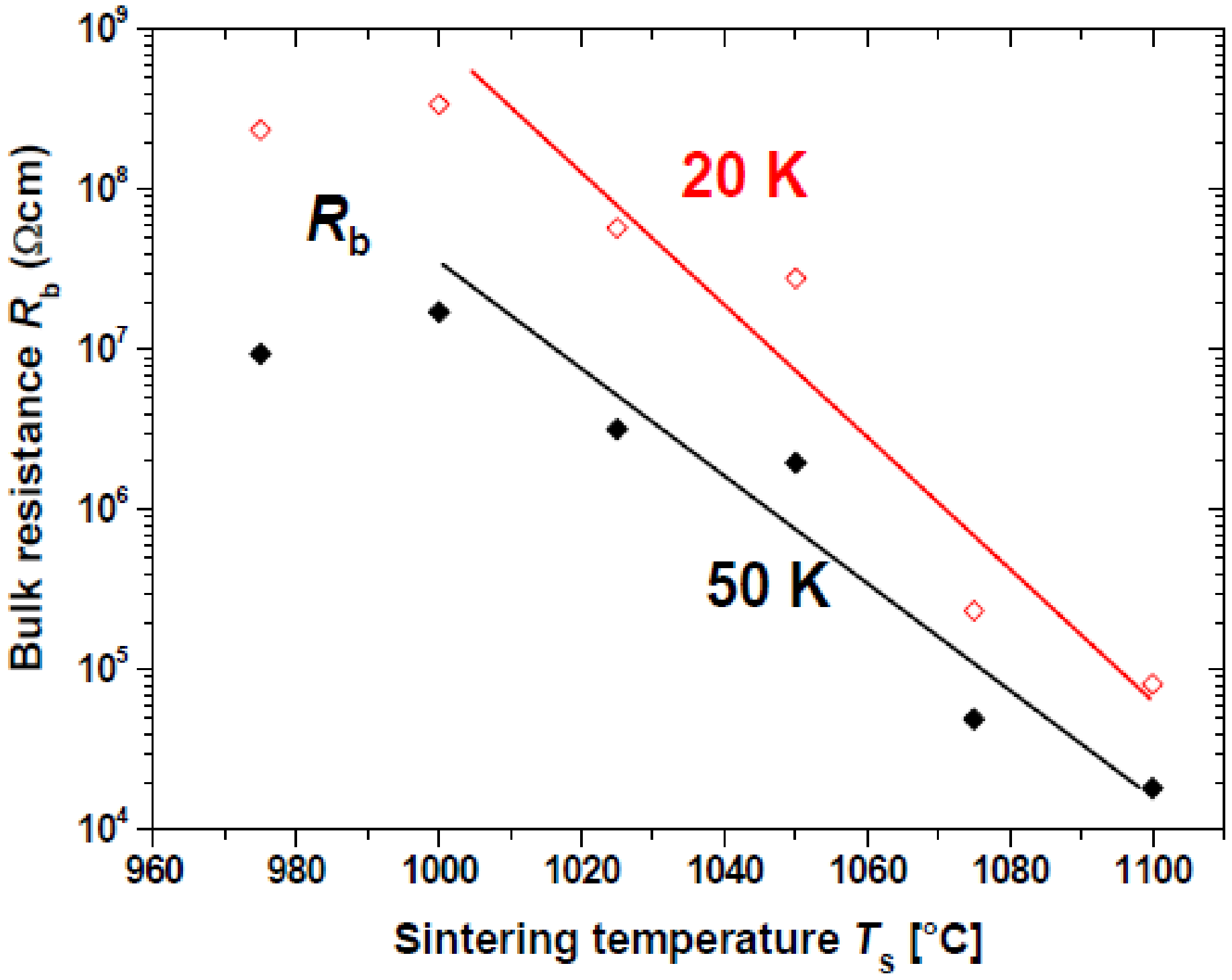

**Figure 17. (a)** $\varepsilon_{GB}$ vs $T_S$ at 120 K. An exponential increase of $\varepsilon_{GB}$ with $T_S$ is shown. **(b)** $\varepsilon_b$ vs $T_S$ at 50 K. A linear increase of $\varepsilon_b$ with $T_S$ is shown. **(c)** $R_b$ vs $T_S$ at 20 (◊) and 50 K (♦). An exponential decrease of $R_b$ with $T_S$ is shown. Solid lines are guide to the eyes. Images reproduced from [52] with permission from the European Ceramic Society.

## C. Influence of Sintering Temperature on the CCTO Composition and Surface Morphology

Scanning electron microscopy (SEM) combined with energy dispersive analysis of X-rays (EDAX) were carried out on all pellets sintered at different $T_S$ to investigate any changes in chemical composition and ceramic microstructure. High resolution quantitative EDAX line scans were carried out using a narrow detector opening to achieve maximum spatial resolution along the single scanning line on the pellet surfaces to demonstrate the development of a Cu-rich segregated inter-granular phase with increasing $T_S$. In an attempt to detect Cu non-stoichiometry in the main CCTO bulk phase, high resolution quantitative EDAX spot analysis on the centre of single CCTO grains was carried out. For each sintered pellet the Ca:Cu:Ti ratios on the centre of 25 different grains were measured to obtain a statistical average for each sample. The penetration depth of the electron beam spot on one single CCTO grain was estimated to be 1 – 2 µm and, therefore, the main X-ray intensity detected was expected to be emitted from the CCTO grain centres irrespective of the inter-granular phase. The quantitative EDAX Ca:Cu:Ti ratios were corrected with respect to standard samples consisting in a CuTi alloy and a $CaTiO_3$ ceramic with predetermined 1:1 Cu:Ti and Ca:Ti atomic ratios. Furthermore, several large area EDAX scans (≈ 150 × 150 µm) with a wide detector opening were carried out on each pellet to obtain the average surface cation concentrations. The mean grain sizes were determined from the respective backscattered SEM images by averaging the size of 220 randomly selected CCTO grains on each sample and the grain size distributions were investigated.

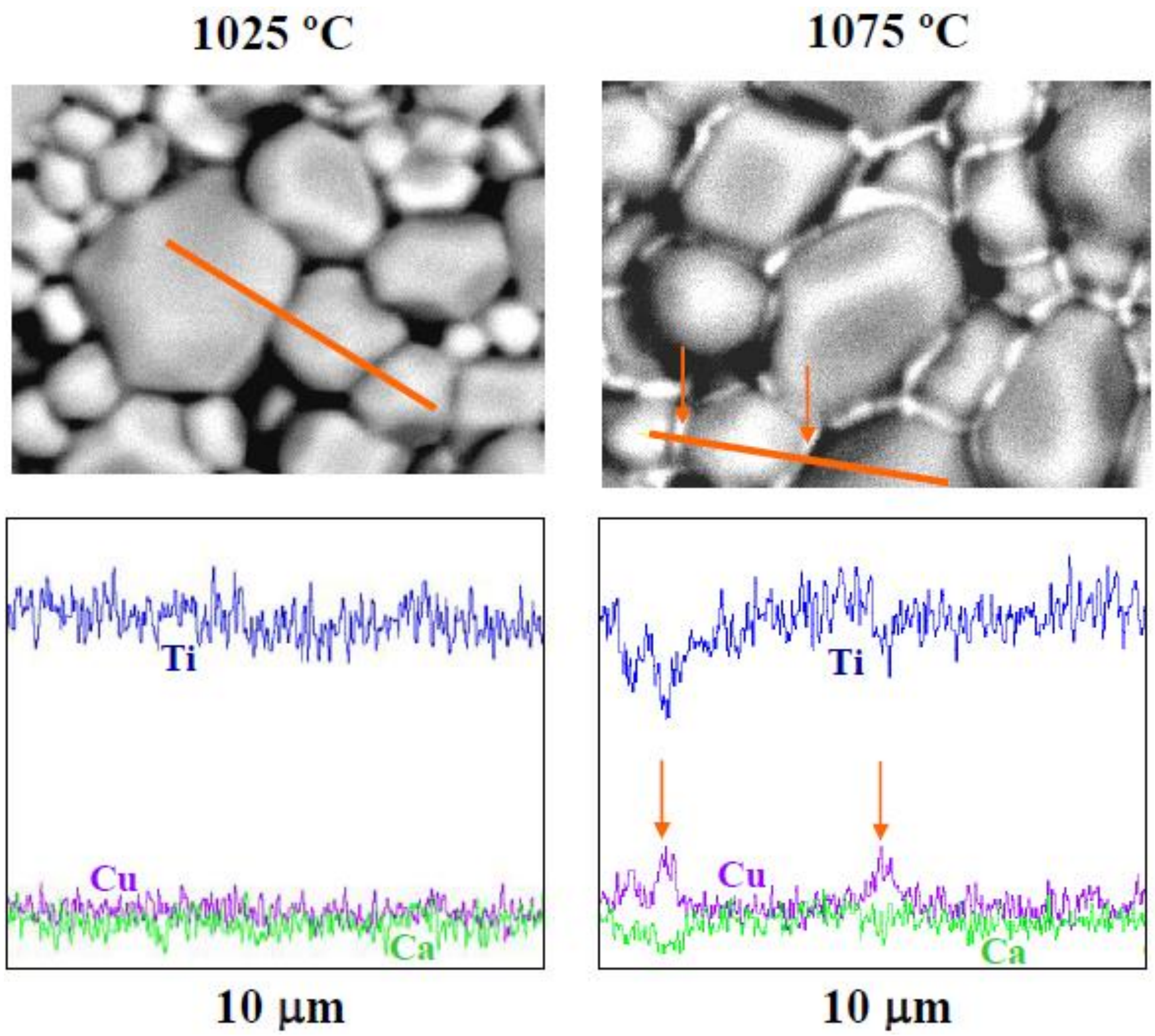

**Figure 18.** SEM images collected in backscattered electron mode from unpolished surfaces of CCTO pellets sintered at 1025 and 1075 °C (first row), and EDAX line scans (second row). The solid (orange) lines in the first row indicate the quantitative EDAX line scans of ≈ 10 µm, with the Ti, Cu, Ca and oxygen atomic concentrations along these lines being presented in the second row. Cu segregation at $T_S$ = 1075 °C in form of an inter-granular phase is indicated. Reproduced from [52] with permission from the European Ceramic Society.

SEM images in the backscattered electron mode and EDAX line scans on the unpolished surfaces of CCTO ceramics sintered at 1025 and 1075 °C are shown in Figure 18. EDAX data were collected from left to right along the solid (false-coloured orange) lines to determine the cation concentrations. In the backscattered mode the image colour contrast emphasizes differences in atomic number and is therefore ideal to detect the segregated Cu-rich secondary phase in CCTO. From this analysis, a Cu-rich phase starts to segregate out of the ceramics at $T_S$ ≈ 1050 °C and accumulates between the CCTO grains as an inter-granular phase. The segregated phase is clearly detectable at 1075 °C whereas there are no detectable signs of it for $T_S$ = 1025 °C. EDAX point measurements for the 1075 °C sintered pellet indicated a high Cu content of the inter-granular phase (about 68 %). The remaining 32% of Ca and Ti cations may well be a proximity effect due to the adjacent CCTO phase and the segregated phase is therefore most likely copper oxide and not a Cu-rich CCTO type phase. This inter-granular $Cu_xO$ phase could not be detected by XRD, probably due to its small volume fraction, or it may be amorphous and no XRD detection was therefore possible. It should be noted that the detected inter-granular phase does not fully cover all CCTO grains and cannot serve as an explanation for the large difference in dielectric properties between GB and bulk relaxations in CCTO. On the other hand, it looks quite likely that GB regions in CCTO would be Cu-rich due to the detected Cu segregation towards the GBs and Cu accumulation as an inter-granular phase.

All XRD patterns collected showed evidence for only the CCTO phase to be present up to $T_S$ = 1100 °C but phase decomposition was indicated above $T_S$ = 1125 °C. For $T_S$ = 1125 °C the XRD pattern showed the presence of $CaTiO_3$, $TiO_2$ and small residues of the main CCTO phase, but no $Cu_xO$ phase. Furthermore, no signs of $Cu_xO$ phase were detected in the SEM backscattered images either after sintering at 1125 °C. This may be understood within the following scenario: $Cu_xO$ segregation starts to occur above $T_S$ = 1050 °C leading to a Cu-deficient CCTO bulk phase and a range of Cu-deficient compositions may exist, constituting one or more defect mechanisms in the form of small solid solutions as mentioned in section III./C./2. Cu segregation occurs towards the GB regions, which may therefore contain a higher Cu-content and display higher resistance in agreement with the arguments proposed in section III./C./2. At higher $T_S$ Cu segregation may intensify and the increasing Cu-deficiency in the CCTO bulk may ultimately lead to the decomposition of the CCTO phase, where the segregated $Cu_xO$ may volatilize. Therefore, the volatility of Cu within the CCTO structure and its segregation towards GBs may be regarded the main driving force for the formation of the IBLC structure in CCTO ceramics.

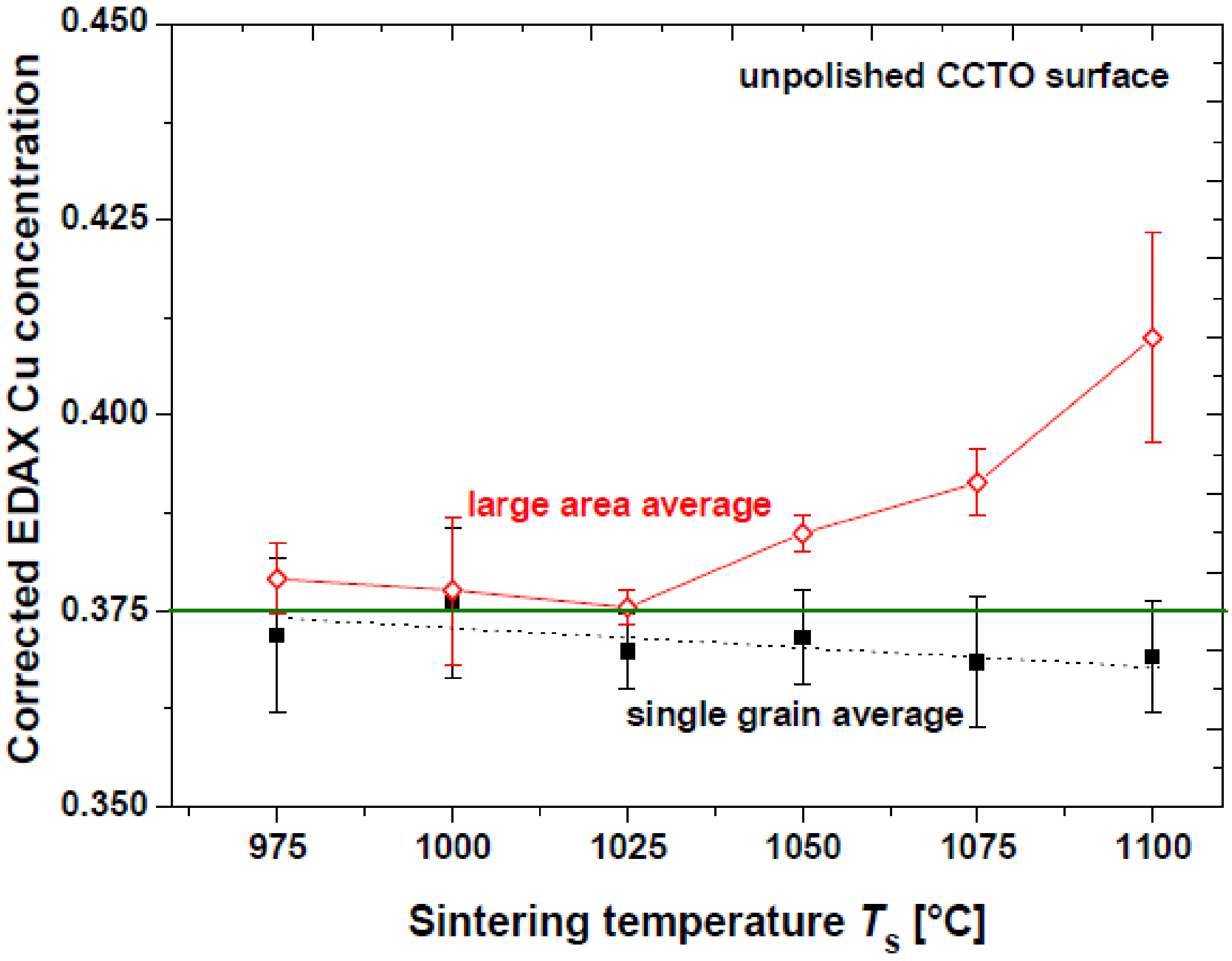

**Figure 19.** Quantitative EDAX results corrected to Ca, Cu and Ti standards. Data were obtained from an average of 25 single grains in each pellet (unpolished) (■) and from large area average scans (5 large scans of 150 × 150 µm for each sample) of the same pellets (◊). Image reproduced from [52] with permission from the European Ceramic Society.

Figure 19 shows the results from high resolution, quantitative EDAX spot analysis of single CCTO ceramic grains, corrected to standard compounds. The Cu contents in terms of the atomic fractions for an average of 25 grains and for the large area average (both for unpolished sintered pellets) are displayed vs $T_S$. The solid horizontal (green) line represents the expected 0.375 Cu cationic ratio (atomic fraction). Cu segregation is clearly evident due to Cu-excess on the unpolished surface above $T_S \geq 1050$ °C. Averaging the Cu content in 25 single grains gave a weak trend towards Cu-loss from within the grain interiors but the trend is statistically just on the limit of significance due to the large experimental error.

The error bars shown represent the first standard deviation of the Cu concentration in the 25 grains investigated, which constitutes a conservative error estimate. Considering that the data points for single grains in Figure 19 represent the highest probability of the actual Cu content of the CCTO grains and that the error estimate is conservative, it does not seem plausible that the clear trend observed on 6 data points is a statistical coincidence. Large area scans taken from polished surfaces showed no Cu accumulation. Obviously, the Cu-rich phase accumulates at the pellet surfaces during sintering and can be removed by polishing. From the backscattered SEM images shown in Figure 20 the mean grain sizes were determined from 220 randomly selected CCTO grains, where the grain size distributions at different $T_S$ shown in Figure 20 indicate a moderate grain growth with increasing $T_S$.

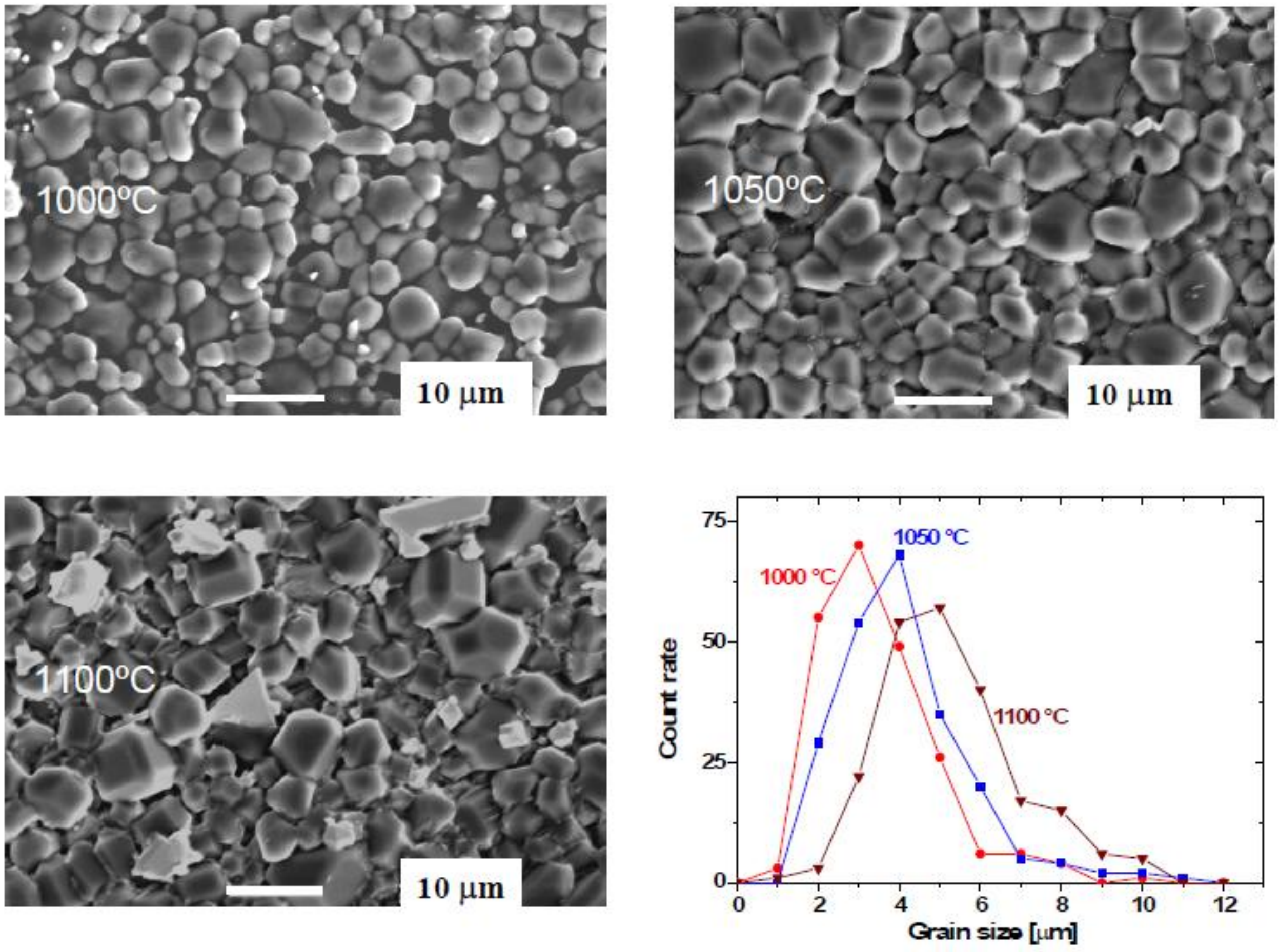

**Figure 20.** SEM images collected with secondary electrons for CCTO pellets sintered at different temperatures, $T_S$. Moderate grain growth is evident with increasing $T_S$. The grain size distributions in the bottom right panel confirm such grain growth.

The average grain size increased approximately linearly from ≈ 3 μm for $T_S$ = 1000 °C to ≈ 4.8 μm for $T_S$ = 1100 °C. The grain growth is accompanied by an increase in pellet density as expected for a densification sintering processes from ≈ 80% for $T_S$ = 1000 °C to ≈ 95% for $T_S$ = 1100 °C. The moderate grain growth detected is insufficient to explain the increase in $\varepsilon_{GB}$ displayed in Figure 14a and 17a as mentioned above.

### D. Summary of the Effects of Sintering Temperature on CCTO Ceramics

Increasing the sintering temperature of CCTO ceramics from 975 to 1100 °C leads to a substantial variation in the dielectric properties. The GB permittivity increases exponentially by a factor of ≈ 300, the bulk permittivity increases linearly by a factor of ≈ 2, and the bulk resistance decreases exponentially by a factor of ≈ $10^3$. The extent of the CCTO defect mechanisms is strongly dependent on $T_S$ and evolves significantly with increasing $T_S$. The bulk semiconductivity is most likely dependent on the Cu content, which was confirmed from quantitative EDAX analysis indicating Cu-loss within individual CCTO grains with increasing $T_S$.

## V. Conclusions and Discussion

CCTO ceramics consist of insulating GB and conducting bulk regions where electronic heterogeneity may be explained by the defect chemistry of the compound. At least two different defect mechanisms exist which may explain (A) resistive GBs and semiconducting bulk by variations in Cu content, and (B) high bulk dielectric permittivity $\varepsilon_b$ in the range of ≈ 100 possibly due to Ca-Cu anti-site defects. The defect mechanisms both develop during the sintering densification process in CCTO ceramics, where variations in Cu content come about due to Cu segregation towards GB regions.

The CCTO giant permittivity values of up to 300 000 reported before are extrinsic in origin, which poses certain limitations for application of CCTO as a capacitor material. Several reports in the literature show that the thin insulating GBs in CCTO exhibit non-linear resistance and may be understood in terms of Schottky-type barriers [3, 16], which is generally disadvantageous in a charge storage device. However, the non-linearity is visible by macroscopic measurement techniques only for CCTO ceramics with very large grain size in the range of 100 μm or more, whereas the CCTO ceramics discussed here showed linear resistance up to the experimental limitation of 20 V. CCTO ceramics with grain sizes in the range of a few micrometers may therefore be good candidates for capacitor applications due to their high permittivity as a result of a core-shell structure of insulating GBs and conducting bulk. Large grain CCTO ceramics on the other hand with strongly non-linear resistance may be interesting to be considered for varistor applications.

## Dedication

The authors wish to dedicate this review to the memory of the late Vito Raineri († 18.2.2012) who led the NUOTO project and is sadly missed by all his colleagues.

## Acknowledgments

The authors wish to acknowledge funding from the European Union (EU) under the NUOTO project and under the FP7/2007-2013 grant agreement nº 226716. R.S. wishes to acknowledge a Ramón y Cajal fellowship from the MICINN/MINECO in Spain.